# New Calculations of the Turbulence-Turbulence Contribution to the Wind Noise Pressure Spectra within Homogeneous Anisotropic Turbulence

Jiao Yu[a,*], Chuanyang Jiang[b,c], Yanying Zhu[a], Jie Wang[a], Cailian Yao[a], Richard Raspet[d], Gregory W. Lyons[d]

[a]College of Science, Liaoning Petrochemical University, Fushun, P. R. China 113001

[b]LTCS and College of Engineering, Peking University, Beijing, P. R. China 100871

[c]Center for Applied Physics and Technology, Peking University, Beijing, P. R. China 100871

[d]National Center for Physical Acoustics, University of Mississippi, MS, USA 38677

The turbulence-turbulence interaction and the turbulence-shear interaction are the sources of intrinsic pressure fluctuation for wind noise generated by atmospheric turbulence. In previous research [Yu et al., J. Acoust. Soc. Am. **129**(2), 622-632 (2011)], it was shown that the measured turbulent fields outdoors can be realistically modeled with Kraichnan's mirror flow model [Kraichnan, J. Acoust. Soc. Am. **28**(3), 378-390 (1956)]. This paper applies Kraichnan's mirror flow idea to develop theory for calculating the turbulence-turbulence interaction wind noise pressure spectra within homogeneous anisotropic turbulence. New calculations of the turbulence-turbulence contribution to the wind noise pressure spectra by incorporating turbulence anisotropy are performed and compared to the result using the same approach but with isotropic input and the result of the turbulence-turbulence interaction pressure spectrum for homogeneous isotropic turbulence using George et al.'s method [George et al., J. Fluid Mech. **148**, 155-191 (1984)]. We also evaluated different contributions to the turbulence-turbulence interaction pressure spectra using our approach with both anisotropic and isotropic inputs. Our results indicate that the turbulence anisotropy has small effect on the turbulence-turbulence interaction pressure in source region, but changes the spectral slope in inertial region to about -5/3. The turbulence-turbulence interaction pressure spectrum incorporating turbulence anisotropy is not sensitive to height. The F33F33 term and the F11F11 term are the most dominant contributions to the anisotropic turbulence pressure spectra in the source region and inertial region, respectively.

[*] Corresponding author. Address: College of Science, Liaoning Petrochemical University, Fushun, 113001, P. R. China. Email: yujiao@lnpu.edu.cn. Tel.: 86-24-56865706. Fax:86-24-56860766.

# I. INTRODUCTION

Wind noise is a significant limitation on acoustic measurements made outdoors under windy conditions. The source of wind noise is locally generated non-propagating pressure fluctuations. The prediction of pressure fluctuations generated in the wind flow above the ground surface outdoors is essential for understanding wind noise in outdoor measurements. When there is no object in the flow, the turbulence-turbulence (t-t) interaction and the turbulence-shear (t-s) interaction are the two contributions to the wind noise. Yu et al. [1] showed that the mirror flow model developed by Kraichnan [2] can be used to model the turbulent fields measured outdoors. The t-s interaction pressure spectrum at the surface was predicted and compared to flush microphone measurements at the surface. Further, based on the work in Yu et al. [1] and by applying the mirror flow model of anisotropic turbulence [2], Yu et al. [3] extended the theory for t-s interaction pressure calculations at the ground surface to the prediction of the t-s interaction pressure spectra at heights above the surface.

The present study is the first known attempt to develop an approach for calculating the t-t interaction pressure spectrum encountered in anisotropic turbulent flow for source and inertial region predictions. There have been extensive efforts devoted to the studies of the pressure fluctuation spectra in isotropic [4-12] and anisotropic [13-19] turbulent flow. Studies to date have been done via direct numerical simulation [8-10,16-19] or experimentally [13-15]. A good portion of literature focuses on studying the pressure spectrum power-law scaling in the inertial range without examining the amplitude. Theories that are well developed on pressure fluctuation spectra are for isotropic turbulence [4-7,20], yet, in contrast, the influence to the pressure spectrum by introducing turbulence anisotropy remains poorly understood due to the difficulty of the computations and the need for a reasonable two-point correlation model. There has been no reported effort to develop theory for the t-t interaction pressure spectrum within homogeneous anisotropic turbulence. This paper attempts to redress this neglect and develop calculations of the t-t interaction pressure spectrum for homogeneous anisotropic turbulence in the source and inertial range.

The works of George et al. [4] and Kraichnan [2] lay a good basis for the present study. George et al. [4] derived explicit forms for the turbulence-turbulence interaction and turbulence-mean shear interaction pressure spectra from velocity spectrum which follows a one dimensional von Karman spectrum form. Their calculations are limited to flow that is incompressible, homogeneous, and isotropic with a constant shear. Kraichnan [2,21] published the first theoretical work investigating pressure fluctuations in boundary layer turbulence. He introduced a mirror flow as an idealized model of the anisotropic boundary layer turbulence, and calculated the total mean square t-s interaction pressure fluctuation on the surface for anisotropic flows. His method will be used to derive the t-t interaction pressure spectral form in anisotropic atmospheric turbulence above the surface.

The remainder of this paper is organized as follows. In Sec. II theory is developed to provide explicit forms for the one-dimensional turbulence-turbulence interaction pressure spectra for the source and inertial region wavenumber space predictions. In Sec. III calculations of the turbulence-turbulence interaction pressure spectra at different heights from given velocity spectral forms are presented. The difference of pressure spectral behaviors from those under isotropic turbulence assumptions is analyzed, and the pressure contributions from different terms are studied. Conclusions are summarized in Sec. IV.

## II. THEORY

We shall consider an incompressible fluid of constant density and viscosity. For a surface in the $x_2$ direction with a mean flow $U_1$ in the $x_1$ direction and velocity fluctuations along all the three directions, the Reynolds-decomposed source equation for the pressure fluctuations is

$$\nabla^2 p = -\rho \left[ 2 \frac{\partial U_1}{\partial x_2} \frac{\partial u_2}{\partial x_1} + \frac{\partial^2}{\partial x_i \partial x_j} (u_i u_j - \overline{u_i u_j}) \right], \qquad (i, j = 1,2,3) \qquad (1)$$

where $p$ is the pressure fluctuation, $u_i$ is the velocity fluctuation in the $x_i$ direction, and $\rho = 1.2 \ kg/m^3$ is the density of air. Note that in Eq. 1, the first term is the turbulence-mean shear interaction term, and the second term is the turbulence-turbulence interaction term.

To model the turbulent field anisotropy in the $x_2$ direction, we define a new velocity field $\tilde{u}_i(\vec{x},t)$ in terms of the fluctuating velocity field $u_i(\vec{x},t)$ of a homogeneous isotropic turbulence:

$$\begin{aligned}
\tilde{u}_1(\vec{x}, t) &= u_1(\vec{x}, t), \\
\tilde{u}_2(\vec{x}, t) &= 2^{-\frac{1}{2}} \left[ u_2(\vec{x}, t) - u_2(\vec{x}^*, t) \right], \\
\tilde{u}_3(\vec{x}, t) &= u_3(\vec{x}, t),
\end{aligned} \qquad (2)$$

where the auxiliary coordinates $\vec{x}^*$ are defined by

$$x_1^* = x_1, x_2^* = -x_2, x_3^* = x_3. \qquad (3)$$

The construction of Eq. 2 is inspired by the mirror flow model developed by Kraichnan [2]. Unlike Kraichnan [2] in which $\tilde{u}_1(\vec{x},t) = 2^{-\frac{1}{2}}[u_1(\vec{x},t) + u_1(\vec{x}^*,t)]$ and $\tilde{u}_3(\vec{x},t) = 2^{-\frac{1}{2}}[u_3(\vec{x},t) + u_3(\vec{x}^*,t)]$ were assumed, we set $\tilde{u}_1(\vec{x},t)$ and $\tilde{u}_3(\vec{x},t)$ to be the same as $u_1(\vec{x}, t)$ and $u_3(\vec{x}, t)$ respectively, to avoid having the surface horizontal components approach twice the amplitude of the components near to the surface. Under the assumption of Eq. 2, the turbulence velocity is homogeneous and isotropic only in the 2D plane parallel to the ground surface. The vertical velocity decreases as it approaches the surface. The normalization by $2^{-\frac{1}{2}}$ yields vertical velocity spectrum identical to the original homogeneous isotropic field when $x_2$ is large.

The source function, $T(\vec{x}, t)$, for the turbulence-turbulence interaction pressure is given by:

$$T(\vec{x}, t) = \rho \frac{\partial^2}{\partial x_i \partial x_j} (\tilde{u}_i \tilde{u}_j - \overline{\tilde{u}_i \tilde{u}_j}), \tag{4}$$

then the correlation function of $T(\vec{x}, t)$ is

$$S(x_2', x_2, x_1' - x_1, x_3' - x_3, t' - t) = \rho^2 \left\langle \frac{\partial^2 (\tilde{u}_i' \tilde{u}_j' - \overline{\tilde{u}_i' \tilde{u}_j'})}{\partial x_i' \partial x_j'} \frac{\partial^2 (\tilde{u}_l \tilde{u}_m - \overline{\tilde{u}_l \tilde{u}_m})}{\partial x_l \partial x_m} \right\rangle$$

$$= \rho^2 \frac{\partial^4 \overline{(\tilde{u}_i' \tilde{u}_j' \tilde{u}_l \tilde{u}_m - \tilde{u}_i' \tilde{u}_j' \overline{\tilde{u}_l \tilde{u}_m} - \overline{\tilde{u}_i' \tilde{u}_j'} \tilde{u}_l \tilde{u}_m + \overline{\tilde{u}_i' \tilde{u}_j'} \, \overline{\tilde{u}_l \tilde{u}_m})}}{\partial x_i' \partial x_j' \partial x_l \partial x_m} \tag{5}$$

$$= \rho^2 \frac{\partial^4 \overline{(u_i' \tilde{u}_j' \tilde{u}_l \tilde{u}_m - \overline{\tilde{u}_i' \tilde{u}_j'} \, \overline{\tilde{u}_l \tilde{u}_m})}}{\partial x_i' \partial x_j' \partial x_l \partial x_m} = \rho^2 \frac{\partial^4 \overline{(\tilde{u}_i' \tilde{u}_l \tilde{u}_j' \tilde{u}_m + \tilde{u}_i' \tilde{u}_m \tilde{u}_j' \tilde{u}_l)}}{\partial x_i' \partial x_l \partial x_j' \partial x_m},$$

where the angular brackets indicate a large-scale ensemble average. In Eq. 5, it is assumed that the fourth-order moments of the velocity are related to the second-order moments in the same way as for a normal joint distribution [5,20].

Letting $\vec{\xi} = \vec{x}' - \vec{x}$ and denoting t'-t by t, we obtain $S(x_2', x_2, \xi_1, \xi_3, t)$ in terms of the double-velocity correlation

$$S(x_2', x_2, \xi_1, \xi_3, t) =$$

$$\rho^2 \{ [2\frac{\partial^4(\tilde{R}_{11}\cdot\tilde{R}_{11})}{\partial\xi_1^4} + 2\frac{\partial^4(\tilde{R}_{33}\cdot\tilde{R}_{33})}{\partial\xi_3^4} + 2\frac{\partial^4(\tilde{R}_{11}\cdot\tilde{R}_{33})}{\partial\xi_1^2\partial\xi_3^2} + 2\frac{\partial^4(\tilde{R}_{33}\cdot\tilde{R}_{11})}{\partial\xi_1^2\partial\xi_3^2}$$

$$+4\frac{\partial^4(\tilde{R}_{11}\cdot\tilde{R}_{13})}{\partial\xi_1^3\partial\xi_3} + 4\frac{\partial^4(\tilde{R}_{13}\cdot\tilde{R}_{11})}{\partial\xi_1^3\partial\xi_3} + 4\frac{\partial^4(\tilde{R}_{33}\cdot\tilde{R}_{13})}{\partial\xi_1\partial\xi_3^3} + 4\frac{\partial^4(\tilde{R}_{13}\cdot\tilde{R}_{33})}{\partial\xi_1\partial\xi_3^3} + 8\frac{\partial^4(\tilde{R}_{13}\cdot\tilde{R}_{13})}{\partial\xi_1^2\partial\xi_3^2}]$$

$$+[2\frac{\partial^4(\tilde{R}_{11}\cdot\tilde{R}_{23})}{\partial x_2'\partial\xi_1^2\partial\xi_3} + 2\frac{\partial^4(\tilde{R}_{23}\cdot\tilde{R}_{11})}{\partial x_2'\partial\xi_1^2\partial\xi_3} - 2\frac{\partial^4(\tilde{R}_{11}\cdot\tilde{R}_{32})}{\partial x_2\partial\xi_1^2\partial\xi_3} - 2\frac{\partial^4(\tilde{R}_{32}\cdot\tilde{R}_{11})}{\partial x_2\partial\xi_1^2\partial\xi_3}$$

$$+2\frac{\partial^4(\tilde{R}_{33}\cdot\tilde{R}_{23})}{\partial x_2'\partial\xi_3^3} + 2\frac{\partial^4(\tilde{R}_{23}\cdot\tilde{R}_{33})}{\partial x_2'\partial\xi_3^3} - 2\frac{\partial^4(\tilde{R}_{33}\cdot\tilde{R}_{32})}{\partial x_2\partial\xi_3^3} - 2\frac{\partial^4(\tilde{R}_{32}\cdot\tilde{R}_{33})}{\partial x_2\partial\xi_3^3}$$

$$+2\frac{\partial^4(\tilde{R}_{11}\cdot\tilde{R}_{21})}{\partial x_2'\partial\xi_1^3} + 2\frac{\partial^4(\tilde{R}_{21}\cdot\tilde{R}_{11})}{\partial x_2'\partial\xi_1^3} - 2\frac{\partial^4(\tilde{R}_{11}\cdot\tilde{R}_{12})}{\partial x_2\partial\xi_1^3} - 2\frac{\partial^4(\tilde{R}_{12}\cdot\tilde{R}_{11})}{\partial x_2\partial\xi_1^3}$$

$$+2\frac{\partial^4(\tilde{R}_{33}\cdot\tilde{R}_{21})}{\partial x_2'\partial\xi_1\partial\xi_3^2} + 2\frac{\partial^4(\tilde{R}_{21}\cdot\tilde{R}_{33})}{\partial x_2'\partial\xi_1\partial\xi_3^2} - 2\frac{\partial^4(\tilde{R}_{33}\cdot\tilde{R}_{12})}{\partial x_2\partial\xi_1\partial\xi_3^2} - 2\frac{\partial^4(\tilde{R}_{12}\cdot\tilde{R}_{33})}{\partial x_2\partial\xi_1\partial\xi_3^2}$$

$$+4\frac{\partial^4(\tilde{R}_{13}\cdot\tilde{R}_{21})}{\partial x_2'\partial\xi_1^2\partial\xi_3} + 4\frac{\partial^4(\tilde{R}_{21}\cdot\tilde{R}_{13})}{\partial x_2'\partial\xi_1^2\partial\xi_3} - 4\frac{\partial^4(\tilde{R}_{13}\cdot\tilde{R}_{12})}{\partial x_2\partial\xi_1^2\partial\xi_3} - 4\frac{\partial^4(\tilde{R}_{12}\cdot\tilde{R}_{13})}{\partial x_2\partial\xi_1^2\partial\xi_3}$$

$$+4\frac{\partial^4(\tilde{R}_{13}\cdot\tilde{R}_{23})}{\partial x_2'\partial\xi_1\partial\xi_3^2} + 4\frac{\partial^4(\tilde{R}_{23}\cdot\tilde{R}_{13})}{\partial x_2'\partial\xi_1\partial\xi_3^2} - 4\frac{\partial^4(\tilde{R}_{13}\cdot\tilde{R}_{32})}{\partial x_2\partial\xi_1\partial\xi_3^2} - 4\frac{\partial^4(\tilde{R}_{32}\cdot\tilde{R}_{13})}{\partial x_2\partial\xi_1\partial\xi_3^2}] \quad (6)$$

$$+[-2\frac{\partial^4(\tilde{R}_{11}\cdot\tilde{R}_{22})}{\partial x_2'\partial x_2\partial\xi_1^2} - 2\frac{\partial^4(\tilde{R}_{22}\cdot\tilde{R}_{11})}{\partial x_2'\partial x_2\partial\xi_1^2} - 2\frac{\partial^4(\tilde{R}_{33}\cdot\tilde{R}_{22})}{\partial x_2'\partial x_2\partial\xi_3^2} - 2\frac{\partial^4(\tilde{R}_{22}\cdot\tilde{R}_{33})}{\partial x_2'\partial x_2\partial\xi_3^2} - 4\frac{\partial^4(\tilde{R}_{13}\cdot\tilde{R}_{22})}{\partial x_2'\partial x_2\partial\xi_1\partial\xi_3}$$

$$-4\frac{\partial^4(\tilde{R}_{22}\cdot\tilde{R}_{13})}{\partial x_2'\partial x_2\partial\xi_1\partial\xi_3} + 2\frac{\partial^4(\tilde{R}_{21}\cdot\tilde{R}_{21})}{\partial x_2'^2\partial\xi_1^2} + 2\frac{\partial^4(\tilde{R}_{12}\cdot\tilde{R}_{12})}{\partial x_2^2\partial\xi_1^2} - 2\frac{\partial^4(\tilde{R}_{12}\cdot\tilde{R}_{21})}{\partial x_2'\partial x_2\partial\xi_1^2} - 2\frac{\partial^4(\tilde{R}_{21}\cdot\tilde{R}_{12})}{\partial x_2'\partial x_2\partial\xi_1^2}$$

$$+2\frac{\partial^4(\tilde{R}_{23}\cdot\tilde{R}_{23})}{\partial x_2'^2\partial\xi_3^2} + 2\frac{\partial^4(\tilde{R}_{32}\cdot\tilde{R}_{32})}{\partial x_2^2\partial\xi_3^2} - 2\frac{\partial^4(\tilde{R}_{32}\cdot\tilde{R}_{23})}{\partial x_2'\partial x_2\partial\xi_3^2} - 2\frac{\partial^4(\tilde{R}_{23}\cdot\tilde{R}_{32})}{\partial x_2'\partial x_2\partial\xi_3^2}$$

$$+2\frac{\partial^4(\tilde{R}_{21}\cdot\tilde{R}_{23})}{\partial x_2'^2\partial\xi_1\partial\xi_3} + 2\frac{\partial^4(\tilde{R}_{12}\cdot\tilde{R}_{32})}{\partial x_2^2\partial\xi_1\partial\xi_3} - 2\frac{\partial^4(\tilde{R}_{12}\cdot\tilde{R}_{23})}{\partial x_2'\partial x_2\partial\xi_1\partial\xi_3} - 2\frac{\partial^4(\tilde{R}_{21}\cdot\tilde{R}_{32})}{\partial x_2'\partial x_2\partial\xi_1\partial\xi_3}$$

$$+2\frac{\partial^4(\tilde{R}_{23}\cdot\tilde{R}_{21})}{\partial x_2'^2\partial\xi_1\partial\xi_3} + 2\frac{\partial^4(\tilde{R}_{32}\cdot\tilde{R}_{12})}{\partial x_2^2\partial\xi_1\partial\xi_3} - 2\frac{\partial^4(\tilde{R}_{32}\cdot\tilde{R}_{21})}{\partial x_2'\partial x_2\partial\xi_1\partial\xi_3} - 2\frac{\partial^4(\tilde{R}_{23}\cdot\tilde{R}_{12})}{\partial x_2'\partial x_2\partial\xi_1\partial\xi_3}]$$

$$+[-2\frac{\partial^4(\tilde{R}_{22}\cdot\tilde{R}_{21})}{\partial x_2'^2\partial x_2\partial\xi_1} + 2\frac{\partial^4(\tilde{R}_{22}\cdot\tilde{R}_{12})}{\partial x_2'\partial x_2^2\partial\xi_1} - 2\frac{\partial^4(\tilde{R}_{21}\cdot\tilde{R}_{22})}{\partial x_2'^2\partial x_2\partial\xi_1} + 2\frac{\partial^4(\tilde{R}_{12}\cdot\tilde{R}_{22})}{\partial x_2'\partial x_2^2\partial\xi_1}$$

$$-2\frac{\partial^4(\tilde{R}_{22}\cdot\tilde{R}_{23})}{\partial x_2'^2\partial x_2\partial\xi_3} + 2\frac{\partial^4(\tilde{R}_{22}\cdot\tilde{R}_{32})}{\partial x_2'\partial x_2^2\partial\xi_3} - 2\frac{\partial^4(\tilde{R}_{23}\cdot\tilde{R}_{22})}{\partial x_2'^2\partial x_2\partial\xi_3} + 2\frac{\partial^4(\tilde{R}_{32}\cdot\tilde{R}_{22})}{\partial x_2'\partial x_2^2\partial\xi_3}] + [2\frac{\partial^4(\tilde{R}_{22}\cdot\tilde{R}_{22})}{\partial x_2'^2\partial x_2^2}]\},$$

where $\tilde{R}_{\alpha\beta}\cdot\tilde{R}_{\mu\upsilon}$ stands for $\tilde{R}_{\alpha\beta}(x_2',x_2,\xi_1,\xi_3,t)\cdot\tilde{R}_{\mu\upsilon}(x_2',x_2,\xi_1,\xi_3,t)$. $\alpha$, $\beta$, $\mu$, $\upsilon$ are introduced to denote each index instead of the summation of the indices. The expression in Eq. 6 is categorized to 5 types with different numbers of vertical velocity contributions in $\alpha$, $\beta$, $\mu$, $\upsilon$, are separated by different square brackets. Note that terms with $\tilde{R}_{13}$ in the numerator and $\partial\xi_1\partial\xi_3$ in the dominator are the same as those with $\tilde{R}_{31}$ in the numerator and $\partial\xi_3\partial\xi_1$ in the dominator. Performing a Fourier transform of Eq. 6 on $\xi_1$, $\xi_3$ and t only,

$$S(x_2',x_2,\vec{\kappa},\omega) = (2\pi)^{-\frac{3}{2}}\int S(x_2',x_2,\xi_1,\xi_3,t)\times\exp(-ik_1\xi_1 - ik_3\xi_3 + i\omega t)d\xi_1 d\xi_3 dt, \quad (7)$$

where $\vec{\kappa}$ is the wave vector in the plane parallel to the boundary, $\vec{\kappa} = k_1 \hat{e}_1 + k_3 \hat{e}_3$, then $S(x_2', x_2, \vec{\kappa}, \omega)$ is given by:

$$
\begin{aligned}
&S(x_2', x_2, \vec{\kappa}, \omega) = \\
&\rho^2 \Big\{ [2k_1^4 F(\tilde{R}_{11} \cdot \tilde{R}_{11}) + 2k_3^4 F(\tilde{R}_{33} \cdot \tilde{R}_{33}) + 2k_1^2 k_3^2 F(\tilde{R}_{11} \cdot \tilde{R}_{33}) + 2k_1^2 k_3^2 F(\tilde{R}_{33} \cdot \tilde{R}_{11}) \\
&+ 4k_1^3 k_3 F(\tilde{R}_{11} \cdot \tilde{R}_{13}) + 4k_1^3 k_3 F(\tilde{R}_{13} \cdot \tilde{R}_{11}) + 4k_1 k_3^3 F(\tilde{R}_{33} \cdot \tilde{R}_{13}) + 4k_1 k_3^3 F(\tilde{R}_{13} \cdot \tilde{R}_{33}) \\
&+ 8k_1^2 k_3^2 F(\tilde{R}_{13} \cdot \tilde{R}_{13})] + \Big[ -2ik_1^2 k_3 \frac{\partial [F(\tilde{R}_{11} \cdot \tilde{R}_{23})]}{\partial x_2'} - 2ik_1^2 k_3 \frac{\partial [F(\tilde{R}_{23} \cdot \tilde{R}_{11})]}{\partial x_2'} \\
&+ 2ik_1^2 k_3 \frac{\partial [F(\tilde{R}_{11} \cdot \tilde{R}_{32})]}{\partial x_2} + 2ik_1^2 k_3 \frac{\partial [F(\tilde{R}_{32} \cdot \tilde{R}_{11})]}{\partial x_2} - 2ik_3^3 \frac{\partial [F(\tilde{R}_{33} \cdot \tilde{R}_{23})]}{\partial x_2'} \\
&- 2ik_3^3 \frac{\partial [F(\tilde{R}_{23} \cdot \tilde{R}_{33})]}{\partial x_2'} + 2ik_3^3 \frac{\partial [F(\tilde{R}_{33} \cdot \tilde{R}_{32})]}{\partial x_2} + 2ik_3^3 \frac{\partial [F(\tilde{R}_{32} \cdot \tilde{R}_{33})]}{\partial x_2} - 2ik_1^3 \frac{\partial [F(\tilde{R}_{11} \cdot \tilde{R}_{21})]}{\partial x_2'} \\
&- 2ik_1^3 \frac{\partial [F(\tilde{R}_{21} \cdot \tilde{R}_{11})]}{\partial x_2'} + 2ik_1^3 \frac{\partial [F(\tilde{R}_{11} \cdot \tilde{R}_{12})]}{\partial x_2} + 2ik_1^3 \frac{\partial [F(\tilde{R}_{12} \cdot \tilde{R}_{11})]}{\partial x_2} - 2ik_1 k_3^2 \frac{\partial [F(\tilde{R}_{33} \cdot \tilde{R}_{21})]}{\partial x_2'} \\
&- 2ik_1 k_3^2 \frac{\partial [F(\tilde{R}_{21} \cdot \tilde{R}_{33})]}{\partial x_2'} + 2ik_1 k_3^2 \frac{\partial [F(\tilde{R}_{33} \cdot \tilde{R}_{12})]}{\partial x_2} + 2ik_1 k_3^2 \frac{\partial [F(\tilde{R}_{12} \cdot \tilde{R}_{33})]}{\partial x_2} \\
&- 4ik_1^2 k_3 \frac{\partial [F(\tilde{R}_{13} \cdot \tilde{R}_{21})]}{\partial x_2'} - 4ik_1^2 k_3 \frac{\partial [F(\tilde{R}_{21} \cdot \tilde{R}_{13})]}{\partial x_2'} + 4ik_1^2 k_3 \frac{\partial [F(\tilde{R}_{13} \cdot \tilde{R}_{12})]}{\partial x_2} \\
&+ 4ik_1^2 k_3 \frac{\partial [F(\tilde{R}_{12} \cdot \tilde{R}_{13})]}{\partial x_2} - 4ik_1 k_3^2 \frac{\partial [F(\tilde{R}_{13} \cdot \tilde{R}_{23})]}{\partial x_2'} - 4ik_1 k_3^2 \frac{\partial [F(\tilde{R}_{23} \cdot \tilde{R}_{13})]}{\partial x_2'} \\
&+ 4ik_1 k_3^2 \frac{\partial [F(\tilde{R}_{13} \cdot \tilde{R}_{32})]}{\partial x_2} + 4ik_1 k_3^2 \frac{\partial [F(\tilde{R}_{32} \cdot \tilde{R}_{13})]}{\partial x_2} \Big] + \Big[ 2k_1^2 \frac{\partial^2 [F(\tilde{R}_{11} \cdot \tilde{R}_{22})]}{\partial x_2' \partial x_2} \\
&+ 2k_1^2 \frac{\partial^2 [F(\tilde{R}_{22} \cdot \tilde{R}_{11})]}{\partial x_2' \partial x_2} + 2k_3^2 \frac{\partial^2 [F(\tilde{R}_{33} \cdot \tilde{R}_{22})]}{\partial x_2' \partial x_2} + 2k_3^2 \frac{\partial^2 [F(\tilde{R}_{22} \cdot \tilde{R}_{33})]}{\partial x_2' \partial x_2} \\
&+ 4k_1 k_3 \frac{\partial^2 [F(\tilde{R}_{13} \cdot \tilde{R}_{22})]}{\partial x_2' \partial x_2} + 4k_1 k_3 \frac{\partial^2 [F(\tilde{R}_{22} \cdot \tilde{R}_{13})]}{\partial x_2' \partial x_2} - 2k_1^2 \frac{\partial^2 [F(\tilde{R}_{21} \cdot \tilde{R}_{21})]}{\partial x_2'^2} \\
&- 2k_1^2 \frac{\partial^2 [F(\tilde{R}_{12} \cdot \tilde{R}_{12})]}{\partial x_2^2} + 2k_1^2 \frac{\partial^2 [F(\tilde{R}_{12} \cdot \tilde{R}_{21})]}{\partial x_2' \partial x_2} + 2k_1^2 \frac{\partial^2 [F(\tilde{R}_{21} \cdot \tilde{R}_{12})]}{\partial x_2 \partial x_2} \\
&- 2k_3^2 \frac{\partial^2 [F(\tilde{R}_{23} \cdot \tilde{R}_{23})]}{\partial x_2'^2} - 2k_3^2 \frac{\partial^2 [F(\tilde{R}_{32} \cdot \tilde{R}_{32})]}{\partial x_2^2} + 2k_3^2 \frac{\partial^2 [F(\tilde{R}_{32} \cdot \tilde{R}_{23})]}{\partial x_2' \partial x_2} \\
&+ 2k_3^2 \frac{\partial^2 [F(\tilde{R}_{23} \cdot \tilde{R}_{32})]}{\partial x_2' \partial x_2} - 2k_1 k_3 \frac{\partial^2 [F(\tilde{R}_{21} \cdot \tilde{R}_{23})]}{\partial x_2'^2} - 2k_1 k_3 \frac{\partial^2 [F(\tilde{R}_{12} \cdot \tilde{R}_{32})]}{\partial x_2^2} \\
&+ 2k_1 k_3 \frac{\partial^2 [F(\tilde{R}_{12} \cdot \tilde{R}_{23})]}{\partial x_2' \partial x_2} + 2k_1 k_3 \frac{\partial^2 [F(\tilde{R}_{21} \cdot \tilde{R}_{32})]}{\partial x_2' \partial x_2} - 2k_1 k_3 \frac{\partial^2 [F(\tilde{R}_{23} \cdot \tilde{R}_{21})]}{\partial x_2'^2} \\
&- 2k_1 k_3 \frac{\partial^2 [F(\tilde{R}_{32} \cdot \tilde{R}_{12})]}{\partial x_2^2} + 2k_1 k_3 \frac{\partial^2 [F(\tilde{R}_{32} \cdot \tilde{R}_{21})]}{\partial x_2' \partial x_2} + 2k_1 k_3 \frac{\partial^2 [F(\tilde{R}_{23} \cdot \tilde{R}_{12})]}{\partial x_2' \partial x_2} \Big] \\
&+ \Big[ -2ik_1 \frac{\partial^3 [F(\tilde{R}_{22} \cdot \tilde{R}_{21})]}{\partial x_2'^2 \partial x_2} + 2ik_1 \frac{\partial^3 [F(\tilde{R}_{22} \cdot \tilde{R}_{12})]}{\partial x_2' \partial x_2^2} - 2ik_1 \frac{\partial^3 [F(\tilde{R}_{21} \cdot \tilde{R}_{22})]}{\partial x_2'^2 \partial x_2} \\
&+ 2ik_1 \frac{\partial^3 [F(\tilde{R}_{12} \cdot \tilde{R}_{22})]}{\partial x_2' \partial x_2^2} - 2ik_3 \frac{\partial^3 [F(\tilde{R}_{22} \cdot \tilde{R}_{23})]}{\partial x_2'^2 \partial x_2} + 2ik_3 \frac{\partial^3 [F(\tilde{R}_{22} \cdot \tilde{R}_{32})]}{\partial x_2' \partial x_2^2} \\
&- 2ik_3 \frac{\partial^3 [F(\tilde{R}_{23} \cdot \tilde{R}_{22})]}{\partial x_2'^2 \partial x_2} + 2ik_3 \frac{\partial^3 [F(\tilde{R}_{32} \cdot \tilde{R}_{22})]}{\partial x_2' \partial x_2^2} \Big] + \Big[ 2 \frac{\partial^4 [F(\tilde{R}_{22} \cdot \tilde{R}_{22})]}{\partial x_2'^2 \partial x_2^2} \Big] \Big\},
\end{aligned} \tag{8}
$$

where $F$ denotes Fourier transform on $\xi_1$, $\xi_3$ and t. The theorem [22] that for any ordinary differentiable functions f(x) and g(x), $\int_{-\infty}^{\infty} f'(x)g(x)dx = -\int_{-\infty}^{\infty} f(x)g'(x)dx$ is applied in Eq. 8. The derivatives on $x_2'$ and $x_2$ can be pulled out of the Fourier transform integral because the Fourier transform is independent of $x_2'$ and $x_2$.

The Fourier transform of the products of the correlation functions can be expressed by a convolution integral as

$$F(\widetilde{R}_{\alpha\beta} \cdot \widetilde{R}_{\mu\nu}) = \frac{1}{(2\pi)^{3/2}} \int \widetilde{R}_{\alpha\beta}(x_2',x_2,\vec{\kappa}-\vec{\kappa}',\omega)\widetilde{R}_{\mu\nu}(x_2',x_2,\vec{\kappa}',\omega)d^2\vec{\kappa}', \qquad (9)$$

which converts the Fourier transform of a product to the convolution form [23].

$\widetilde{R}_{\alpha\beta}(x_2',x_2,\vec{\kappa},\omega)$ may be expressed as:

$$\begin{aligned}
&\widetilde{R}_{11}(x_2',x_2,\vec{\kappa},\omega) = R_{11}(x_2'-x_2,\vec{\kappa},\omega), \quad \widetilde{R}_{33}(x_2',x_2,\vec{\kappa},\omega) = R_{33}(x_2'-x_2,\vec{\kappa},\omega), \\
&\widetilde{R}_{13}(x_2',x_2,\vec{\kappa},\omega) = R_{13}(x_2'-x_2,\vec{\kappa},\omega), \quad \widetilde{R}_{31}(x_2',x_2,\vec{\kappa},\omega) = R_{31}(x_2'-x_2,\vec{\kappa},\omega), \\
&\widetilde{R}_{21}(x_2',x_2,\vec{\kappa},\omega) = \frac{1}{\sqrt{2}}[R_{21}(x_2'-x_2,\vec{\kappa},\omega) - R_{21}(-x_2'-x_2,\vec{\kappa},\omega)], \\
&\widetilde{R}_{12}(x_2',x_2,\vec{\kappa},\omega) = \frac{1}{\sqrt{2}}[R_{12}(x_2'-x_2,\vec{\kappa},\omega) - R_{12}(x_2'+x_2,\vec{\kappa},\omega)], \\
&\widetilde{R}_{23}(x_2',x_2,\vec{\kappa},\omega) = \frac{1}{\sqrt{2}}[R_{23}(x_2'-x_2,\vec{\kappa},\omega) - R_{23}(-x_2'-x_2,\vec{\kappa},\omega)], \\
&\widetilde{R}_{32}(x_2',x_2,\vec{\kappa},\omega) = \frac{1}{\sqrt{2}}[R_{32}(x_2'-x_2,\vec{\kappa},\omega) - R_{32}(x_2'+x_2,\vec{\kappa},\omega)], \\
&\widetilde{R}_{22}(x_2',x_2,\vec{\kappa},\omega) = \Re_{22}(x_2'-x_2,\vec{\kappa},\omega) - \Re_{22}(x_2'+x_2,\vec{\kappa},\omega),
\end{aligned} \qquad (10)$$

where $\Re_{22}(x_2,\vec{\kappa},\omega)$ is the real part of the Fourier transform $R_{22}(x_2,\vec{\kappa},\omega)$ [2]. Using Eq. 10, we can derive $F(\widetilde{R}_{\alpha\beta} \cdot \widetilde{R}_{\mu\nu})$ in terms of the spectrum function of the homogeneous turbulence. For $(\widetilde{R}_{\alpha\beta} \cdot \widetilde{R}_{\mu\nu}) = (\widetilde{R}_{11} \cdot \widetilde{R}_{11}), (\widetilde{R}_{33} \cdot \widetilde{R}_{33}), (\widetilde{R}_{11} \cdot \widetilde{R}_{33}), (\widetilde{R}_{33} \cdot \widetilde{R}_{11}), (\widetilde{R}_{11} \cdot \widetilde{R}_{13}), (\widetilde{R}_{13} \cdot \widetilde{R}_{11}), (\widetilde{R}_{33} \cdot \widetilde{R}_{13}), (\widetilde{R}_{13} \cdot \widetilde{R}_{33}), (\widetilde{R}_{13} \cdot \widetilde{R}_{13})$,

$$\begin{aligned}
F(\widetilde{R}_{\alpha\beta} \cdot \widetilde{R}_{\mu\nu}) &= \frac{1}{(2\pi)^{3/2}} \int R_{\alpha\beta}(x_2'-x_2,\vec{\kappa}-\vec{\kappa}')R_{\mu\nu}(x_2'-x_2,\vec{\kappa}')d^2\vec{\kappa}' \\
&= \frac{1}{(2\pi)^{3/2}} \int \left(\frac{1}{\sqrt{2\pi}} \int R_{\alpha\beta}(\vec{k}-\vec{k}') \cdot e^{i(k_2-k_2')(x_2'-x_2)}d(k_2-k_2')\right)\left(\frac{1}{\sqrt{2\pi}} \int R_{\mu\nu}(\vec{k}') \cdot e^{ik_2'(x_2'-x_2)}dk_2'\right)d^2\vec{\kappa}' \quad (11) \\
&= \frac{1}{(2\pi)^{5/2}} \int e^{ik_2(x_2'-x_2)}\left(\int R_{\alpha\beta}(\vec{k}-\vec{k}')R_{\mu\nu}(\vec{k}')d^3\vec{k}'\right)dk_2,
\end{aligned}$$

where $R_{\alpha\beta}(\vec{k})$ (or $R_{\mu\nu}(\vec{k})$) is the spectrum function of the homogeneous turbulence. All other terms in Eq. 8, $F(\widetilde{R}_{\alpha\beta} \cdot \widetilde{R}_{\mu\nu})$ may be converted in similar ways (Appendix A). In Eq. 11 and Appendix A, for simplicity, $\omega$ is omitted in the form. $R_{\alpha\beta}(x_2'-x_2,\vec{\kappa}-\vec{\kappa}')R_{\mu\nu}(x_2'-x_2,\vec{\kappa}')$ stands for $R_{\alpha\beta}(x_2'-x_2,\vec{\kappa}-\vec{\kappa}',\omega)R_{\mu\nu}(x_2'-x_2,\vec{\kappa}',\omega)$ and $R_{\alpha\beta}(\vec{k}-\vec{k}')R_{\mu\nu}(\vec{k}')$ stands for $R_{\alpha\beta}(\vec{k}-\vec{k}',\omega)R_{\mu\nu}(\vec{k}',\omega)$. The derivatives of $F(\tilde{R}_{\alpha\beta}\cdot\tilde{R}_{\mu\nu})$ with respect to $x_2'$ and $x_2$ as in Eq. 8 are easily derived.

For the pressure calculation in the half-space with rigid surface at $x_2=0$ and bounded as $x_2$ approaches infinity, Yu et al. [3] derived the power spectrum form of the pressure transform normalized to unit area and time as a function of height:

$$|p(x_2,\vec{\kappa},\omega)|^2 = \frac{1}{4}(2\pi)^{-\frac{3}{2}}\kappa^{-2}[\int_0^\infty\int_0^\infty e^{-\kappa|x_2'-x_2|-\kappa|x_2''-x_2|}S(x_2'',x_2',\vec{\kappa},\omega)dx_2'dx_2''$$
$$+e^{-\kappa x_2}\int_0^\infty\int_0^\infty e^{-\kappa x_2''-\kappa|x_2'-x_2|}S(x_2'',x_2',\vec{\kappa},\omega)dx_2'dx_2''$$
$$+e^{-\kappa x_2}\int_0^\infty\int_0^\infty e^{-\kappa x_2'-\kappa|x_2''-x_2|}S(x_2'',x_2',\vec{\kappa},\omega)dx_2'dx_2''$$
$$+e^{-2\kappa x_2}\int_0^\infty\int_0^\infty e^{-\kappa x_2'-\kappa x_2''}S(x_2'',x_2',\vec{\kappa},\omega)dx_2'dx_2''].$$
(12)

When calculating the turbulence-shear interaction pressure spectrum in Yu et al. [3], we used the Fourier transform of the turbulence-shear interaction source function correlation tensor form for $S(x_2'',x_2',\vec{\kappa},\omega)$. Here Eq. 12 is valid, but $S(x_2',x_2,\vec{\kappa},\omega)$ in Eq. 8 should be used, except that $x_2'$ and $x_2$ used in previous circumstances should be changed to $x_2''$ and $x_2'$ respectively, to avoid confusion.

$S(x_2'',x_2',\vec{\kappa},\omega)$ in Eq. 8 is the summation of many terms with the form $\int k_2^s e^{\pm ik_2 G(x_2'',x_2')}\left(\int k_2'^t e^{\pm 2ik_2'H(x_2'',x_2')}R_{\alpha\beta}(\vec{k}-\vec{k}')R_{\mu\nu}(\vec{k}')d^3\vec{k}'\right)dk_2$, where $s$, $t$ are power indices with the value 0-4, $\omega$ is omitted for simplicity, and $G(x_2'',x_2')$, $H(x_2'',x_2')$ may be $x_2''-x_2'$, $x_2''+x_2'$, $x_2''$, $x_2'$, or 0. The two $\pm$ signs mean positive or negative, and the signs in either exponential are chosen independently. $k_2^s$ and $k_2'^t$ are generated by taking the derivatives of the terms in Appendix A as in Eq. 8. Note that the imaginary unit $i$ in Eq. 8 will be eliminated after the derivative calculation.

Switching the order of integrations in Eq. 12, then integrating $x_2''$ and $x_2'$ first, will eliminate $x_2''$ and $x_2'$ and leave only $x_2$, $\vec{k}$, $\vec{k}'$, $\omega$ in the expression of Eq. 12 with the integrations on $d^3\vec{k}'$ and $dk_2$.

Ignoring $k_2^s$ and $k_2'^t$, $S(x_2'',x_2',\vec{\kappa},\omega)$ is the summation of terms with $\int e^{\pm ik_2 G(x_2'',x_2')}\left(\int e^{\pm 2ik_2'H(x_2'',x_2')}R_{\alpha\beta}(\vec{k}-\vec{k}')R_{\mu\nu}(\vec{k}')d^3\vec{k}'\right)dk_2$ for 16 cases, which may be found from Eq. 11 and Appendix A. First, we analyze each of the 16 cases, do integrations on $x_2''$ and $x_2'$ following procedures in the square bracket in Eq. 12 for each case (Appendix B). Then we reinsert the $k_2^s$ and $k_2'^t$ terms, together with the coefficients, and sum all terms together to obtain the form for $|p(x_2, \vec{\kappa}, \omega)|^2$. To calculate the integrations, the integrations are factored to two regions $[0, x_2]$ and $[x_2, \infty]$ for both integration variables $x_2''$ and $x_2'$ to derive the results shown in Appendix B. In Appendix B, $\omega$ is also omitted for simplicity. Note that $R_{\alpha\beta}(\vec{k}-\vec{k}')R_{\mu\nu}(\vec{k}')$ is equivalent to $R_{\beta\alpha}(\vec{k}-\vec{k}')R_{\mu\nu}(\vec{k}')$ or $R_{\alpha\beta}(\vec{k}-\vec{k}')R_{\nu\mu}(\vec{k}')$ or $R_{\beta\alpha}(\vec{k}-\vec{k}')R_{\mu\nu}(\vec{k}')$.

Integrating $|p(x_2, \vec{\kappa}, \omega)|^2$ over $\omega$ gives $|p(x_2, \vec{\kappa})|^2$. Note that

$$\int_{-\infty}^{\infty} R_{\alpha\beta}(\vec{k}-\vec{k}', \omega) R_{\mu\nu}(\vec{k}', \omega) d\omega = 16\pi^4 F_{\alpha\beta}(\vec{k}-\vec{k}') F_{\mu\nu}(\vec{k}'), \quad (13)$$

where $F_{\alpha\beta}(\vec{k})$ is the 3D energy spectrum tensor for homogenous turbulence defined by Batchelor [20]. (The numerical factor arises from a difference in normalization of the Fourier transform [2]). $F_{\alpha\beta}(\vec{k})$ satisfies

$$F_{\alpha\beta}(\vec{k}) = \frac{E(k)[k^2 \delta_{\alpha\beta} - k_\alpha k_\beta]}{4\pi k^4}, \quad (14)$$

where $E(k)$ is the three dimensional velocity spectrum function. In Reference 12, it is shown that the measured longitudinal velocity spectrum above the surface outdoors can be fit to a revised von Karman form

$$F_{11}^1(k_1) = \frac{C}{[1+(k_1\lambda)^2]^{5/6}}, \quad (15)$$

where $k_1$ is the wave number in the direction of flow, and the $C$ and $\lambda$ are the fit parameters. The form for $E(k)$ corresponding to Eq. 15 is given by [4,12]

$$E(k) = \frac{55C}{18} \frac{(k\lambda)^4}{[1+(k\lambda)^2]^{17/6}}. \quad (16)$$

Integrating $|p(x_2, \vec{\kappa})|^2$ over $k_3$ gives the turbulence-turbulence interaction pressure fluctuation spectrum $|p(x_2, k_1)|^2$ at height $x_2$ (Appendix C). To convert double-sided spectrum to one-sided spectrum with only positive $k_1$, a factor of two is introduced. The form of $|p(x_2, k_1)|^2$ (Appendix C) is the desired form for comparison with pressure spectral measurements under Taylor's hypothesis.

## III. CALCULATION AND DISCUSSION

Section III A presents calculations of the turbulence-turbulence interaction pressure spectra at different heights from given velocity spectral forms, and compares the pressure spectral behaviors under anisotropic and isotropic turbulence assumptions. Section III B studies the pressure contributions from different terms under anisotropic and isotropic turbulence assumptions.

**A. Calculations of the t-t interaction pressure spectra at different heights and comparisons with pressure spectra under isotropic turbulence assumptions**

The expression of $|p(x_2,k_1)|^2$ presented in Appendix C is evaluated with MATHEMATICA© (Wolfram Research, http://www.wolfram.com) for two runs: Run 1 and Run 2. Table 1 lists the parameters used in the calculations for the two runs. The C and $\lambda$ values for Run 1 and Run 2 are obtained from Ref. 3 by fitting the measured longitudinal wind velocity spectra at 1 meter height above the ground outdoors to the revised von Karman form in Eq. 15.

For comparisons with the anisotropic results, results using the same method but with homogeneous isotropic input in Eq. 2 and results of the turbulence-turbulence interaction pressure spectrum for homogeneous isotropic turbulence using George et al.'s method [4] following Ref. 12 are also displayed. Appendix D lists the expression (Eq. D-3) used for calculations using the same method but with homogeneous isotropic input. Eq. 14 in Ref. 12 is the expression for homogeneous isotropic turbulence t-t interaction pressure spectral calculations using George et al.'s method [4].

**Table 1. Parameters used in calculations.**

| Run | C | $\lambda$ |
|---|---|---|
| 1 | 2.35 | 4.34 |
| 2 | 7.60 | 11.30 |

Figure 1 displays the velocity spectrum fits and the predicted turbulence-turbulence interaction pressure fluctuation spectra. Figures 1(a) and 1(b) display velocity spectrum fits for Run 1 and Run 2 respectively. Figures 1(c) and 1(d) display the predicted t-t interaction pressure fluctuation spectra for Run 1 and Run 2. In Figs. 1(c) and 1(d), the t-t interaction pressure spectra using the same method but with homogeneous isotropic turbulence input and using George et al.'s method following Ref. 12 for homogeneous isotropic turbulence are calculated for comparison with the anisotropic results.

From Figs. 1(a), 1(b), it can be seen that the velocity spectra follow a -5/3 law in high wave numbers, and the inertial region starts from a lower wave number for the fit with a larger $\lambda$. In both Figs. 1(c) and 1(d), predicted pressure spectra for anisotropic turbulence level off at low wave numbers and present a spectral slope about (slightly steeper than) -5/3 in the inertial region. Under anisotropic turbulence condition, the turbulence-turbulence interaction pressure fluctuation spectra do not satisfy the exact power-law relation like in the isotropic condition in the logarithmic coordinate but exhibit curls in the shape.

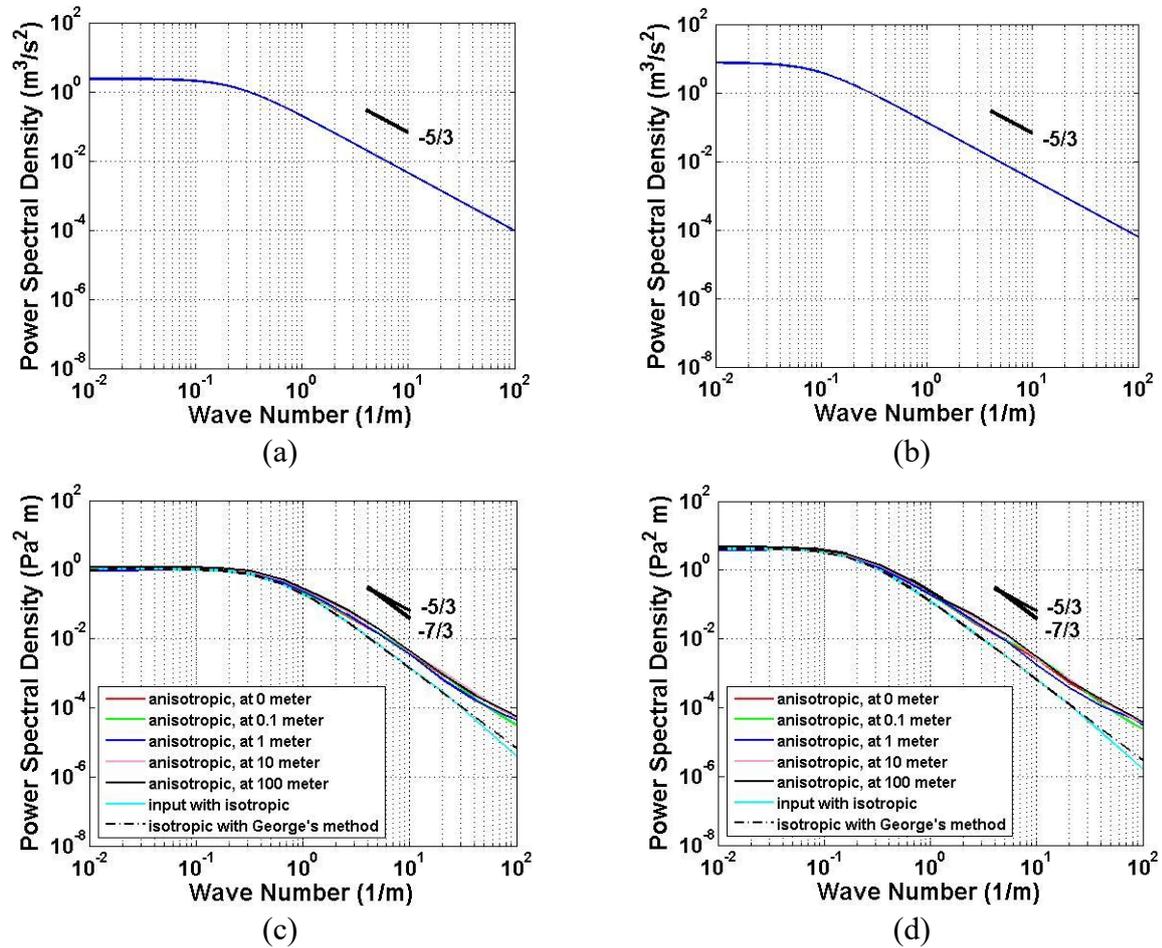

**Figure 1. Predefined velocity spectrum fits (a), (b) and predicted turbulence-turbulence interaction pressure fluctuation spectra (c), (d). (a) Run 1 predefined velocity spectrum fit. (b) Run 2 predefined velocity spectrum fit. (c) Run 1 predicted t-t interaction pressure fluctuation spectrum. (d) Run 2 predicted t-t interaction pressure fluctuation spectrum.**

It can be clearly seen that for the isotropic input calculations using Eq. D-3, the pressure spectra match up with anisotropic results at low wave number, but gradually deviate with a -7/3 slope in the inertial region as wave number increases. The plot of calculation with the isotropic input using Eq. D-3 overlaps with the plot of calculation with George's method using Eq. 14 in Ref. 12 in both the source region and inertial region which provides strong support to the justification of our method. The discrepancy for wave number greater than 40 m-1 is probably numerical error. For the anisotropic turbulence in this study, predicted t-t interaction pressure spectra at different heights only differ slightly which indicates that the turbulence-turbulence interaction pressure spectrum is not sensitive to height.

The trends and levels of the pressure spectra in Fig. 1(d) compared to those in Fig. 1(c) are consistent with the velocity spectrum fit in Fig. 1(b) relative to that in Fig. 1(a). In the pressure spectrum predictions, the wave number from where the source region transitions to the inertial region still depends on $\lambda$; larger $\lambda$ value corresponds to lower wave number.

## B. The pressure contributions from different terms under anisotropic and isotropic turbulence assumptions

In order to study the pressure spectrum comparison results between anisotropic turbulence and isotropic turbulence input, contributions from different terms in Appendix C (Eq. C-1) and Appendix D (Eq. D-3) are calculated for Run 1. Figure 2 displays different contributions to Run 1 predicted t-t interaction pressure fluctuation spectra.

From Fig. 2, it can be seen that at low wave number, the largest contribution for anisotropic turbulence calculation comes from the $F_{33}F_{33}$ term, and the largest two contributions for isotropic turbulence input calculation come from the $F_{33}F_{33}$ term and the $F_{22}F_{22}$ term, all of which level off at low wave number. In inertial region, the largest contribution comes from the $F_{11}F_{11}$ term which falls off with a -5/3 rate, for both anisotropic and isotropic turbulence input.

In Fig. 2(a), for the same $F_{\alpha\beta}F_{\mu\nu}$ contribution, all anisotropic predictions are higher than the corresponding isotropic predictions with similar trends at low wave number, but the anisotropic and isotropic predictions join together in inertial region. Therefore, the terms in Fig. 2(a) do not contribute to the deviation of the anisotropic turbulence prediction from the isotropic turbulence input prediction at high wave number, but Fig. 2(a) displays a higher value for anisotropic prediction at low wave number compared to isotropic prediction.

Figure 2(b) displays contributions from terms with three or four vertical velocity components in $F_{\alpha\beta}(\vec{k}-\vec{k}')F_{\mu\nu}(\vec{k}')$. For all the terms in Fig. 2(b), anisotropic turbulence prediction lines are lower than the corresponding isotropic turbulence prediction lines in magnitude (with less positive/negative values) in the full wave number region. For wave numbers greater than $1 m^{-1}$, the $F_{22}F_{22}$ anisotropic and isotropic turbulence prediction lines are parallel. But all the other terms' anisotropic turbulence prediction lines drop off more rapidly than the corresponding isotropic turbulence predictions as wave number increases. Considering that they all have negative values, all these terms contribute to the high wave number deviation from the -7/3 scaling and the higher value of anisotropic prediction than the isotropic in inertial region in Fig. 1(c).

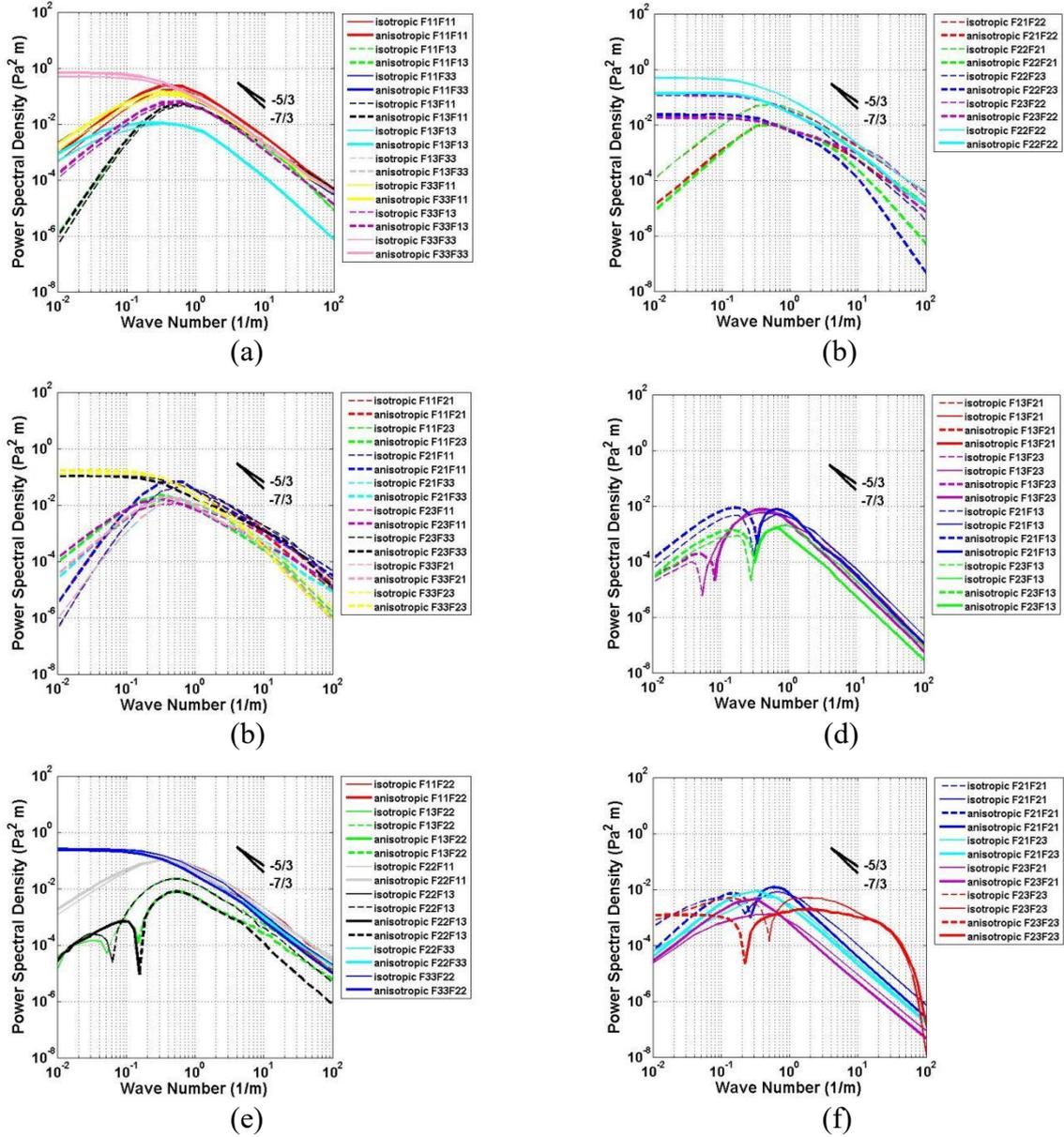

Figure 2. Different contributions to Run 1 predicted t-t interaction pressure fluctuation spectra. Lines labeled anisotropic are contributions to pressure spectrum for anisotropic turbulence at 1 meter. Lines labeled isotropic are contributions to pressure spectrum for calculation with isotropic turbulence input. Solid lines are positive, and dashed lines are negative. Anisotropic lines are made thicker than isotropic lines for better distinction. Six different images are contributions from terms with different numbers of vertical velocity components in $F_{\alpha\beta}(\vec{k}-\vec{k}')F_{\mu\nu}(\vec{k}')$: (a) no vertical velocity component; (b) three and four vertical velocity components; (c) one vertical velocity component; (d) one vertical velocity component (continued); (e) two vertical velocity components; (f) two vertical velocity components (continued).

All the contributions in Fig. 2(c) have negative values. Anisotropic predictions are more negative than isotropic predictions in source region, and less negative than isotropic predictions in high wave number region. All the terms in Fig. 2(c) contribute to the larger total value of anisotropic prediction than the isotropic in Fig. 1(c) at high wave number. The contributions from Fig. 2(d) are negligible. All the contributions in Fig. 2(d) are first negative then positive as wave number increases. This sudden transition of the power spectra from negative to positive is intriguing. The negative fluctuations correspond to low-pressure regions in space. It is known [8,24] that the high-vorticity regions are strongly correlated with the low-pressure regions. It may be the high vorticity that are corresponding to the negative spectra in wavenumber space.

At low wave number in Fig. 2(e), the anisotropic and isotropic $F_{33}F_{22}$ lines overlap with the anisotropic and isotropic $F_{22}F_{33}$ lines respectively, and the anisotropic and isotropic $F_{22}F_{11}$ lines overlap with the anisotropic and isotropic $F_{11}F_{22}$ lines respectively. At high wave number, all anisotropic turbulence lines have lower magnitude than the corresponding isotropic turbulence lines, for both positive and negative values. Lines with positive values contribute negatively to the total contribution comparison result in Fig. 1(c) at high wave number, and lines with negative values contribute positively to the total comparison result in Fig. 1(c). In Fig. 2(f), anisotropic and isotropic $F_{23}F_{23}$ lines contribute most at high wave number. The rest of the lines display lower values for anisotropic predictions than the corresponding isotropic predictions, with the values all positive at high wave number, but their spectral levels are relatively much smaller, and hence have a very little effect to the total result.

Table 2 lists the main contributions to Run 1 t-t interaction pressure fluctuation spectra at $k_1 = 10^{-2}\ m^{-1}$ for anisotropic turbulence calculation at 1 meter and calculation with isotropic turbulence input. Contributions with absolute values below $10^{-2}\ Pa^2\ m$ are not listed. For reference, the total results in Fig. 1(c) at $k_1 = 10^{-2}\ m^{-1}$ for anisotropic turbulence calculation at 1 meter，calculation with isotropic turbulence input and calculation using George et al.'s method for homogeneous isotropic turbulence are listed.

From Table 2, it can be seen that none of the main contributions to the low wave number pressure spectra have any longitudinal velocity component in $F_{\alpha\beta}F_{\mu\nu}$. The largest two contributions for isotropic turbulence input calculation--- the $F_{33}F_{33}$ term and the $F_{22}F_{22}$ term have equal contributions. Similarly, for isotropic turbulence prediction, there are other equal contributions, such as the $F_{33}F_{22}$ term and the $F_{22}F_{33}$ term, etc. Summing up all the main contributions in Table 2 gives a value of 0.98 for anisotropic turbulence calculation and a value of 1.03 for isotropic turbulence input calculation. The differences between the two values and the corresponding total values in Fig. 1(c) as listed in Table 2 are caused by the terms not listed whose contributions are very small. The values for calculation with isotropic turbulence input and calculation using George et al.'s method for homogeneous isotropic turbulence at $k_1 = 10^{-2} m^{-1}$ in Fig. 1(c) are identical (=1.04 $Pa^2 m$). It is clear that at low wave number, anisotropic and isotropic turbulence predicted spectra have very close values, despite that they have different dominant terms.

**Table 2. Main contributions to Run 1 t-t interaction pressure fluctuation spectra at $k_1 = 10^{-2} m^{-1}$ for anisotropic turbulence calculation at 1 meter and calculation with isotropic turbulence input.**[*]

| $k_1 = 10^{-2} m^{-1}$ | $F_{33}F_{33}$ ($Pa^2 m$) | $F_{22}F_{22}$ ($Pa^2 m$) | $F_{33}F_{22}$ ($Pa^2 m$) | $F_{22}F_{33}$ ($Pa^2 m$) | $F_{23}F_{22}$ ($Pa^2 m$) | $F_{22}F_{23}$ ($Pa^2 m$) | $F_{33}F_{23}$ ($Pa^2 m$) | $F_{23}F_{33}$ ($Pa^2 m$) | Total in Fig. 1(c) ($Pa^2 m$) |
|---|---|---|---|---|---|---|---|---|---|
| Anisotropic @1 m | 0.692 | 0.134 | 0.233 | 0.241 | -0.018 | -0.023 | -0.168 | -0.110 | 0.95 |
| Isotropic input | 0.490 | 0.490 | 0.261 | 0.261 | -0.113 | -0.123 | -0.123 | -0.113 | 1.04 |
| Isotropic (George) | - | - | - | - | - | - | - | - | 1.04 |

[*]Note: Contributions with absolute values below $10^{-2} Pa^2 m$ are not listed.

Table 3 lists the main contributions to Run 1 t-t interaction pressure fluctuation spectra at $k_1 = 10 m^{-1}$ for anisotropic turbulence calculation at 1 meter and calculation with isotropic turbulence input. Contributions with absolute values below $10^{-4} Pa^2 m$ are not listed. Table 3 is categorized to four stories based on the different numbers of vertical velocity contributions in $\alpha$, $\beta$, $\mu$, $\upsilon$ of $F_{\alpha\beta}F_{\mu\nu}$. For reference, the total results in Fig. 1(c) at $k_1 = 10 m^{-1}$ for anisotropic turbulence calculation at 1 meter, calculation with isotropic turbulence input and calculation using George et al.'s method for homogeneous isotropic turbulence are listed.

From Table 3, it can be seen that for terms with no vertical velocity component in $F_{\alpha\beta}F_{\mu\nu}$, the anisotropic and isotropic turbulence input calculations have equal values in the inertial region. For all the rest terms, the anisotropic turbulence calculations are lower in magnitude than the corresponding isotropic turbulence calculations, for both negative and positive values. In other words, the anisotropic turbulence calculation results are less negative and less positive than their isotropic counterparts. Summing up all the main contributions in Table 3 gives a value of $2.89 \times 10^{-3} Pa^2 m$ for anisotropic turbulence calculation and a value of $0.94 \times 10^{-3} Pa^2 m$ for isotropic turbulence input calculation. Their differences from the corresponding total values in Fig. 1(c) as listed in Table 3 are caused by the terms not listed. The values for calculation with isotropic turbulence input and calculation using George et al.'s method for homogeneous isotropic turbulence at $k_1 = 10 m^{-1}$ in Fig. 1(c) are identical (=$1.33 \times 10^{-3} Pa^2 m$). At high wave number, anisotropic turbulence predicted spectrum is higher than the isotropic turbulence calculated spectrum. The summation is an outcome of less negative amount being larger than the less positive amount for the anisotropic turbulence prediction compared to isotropic. For isotropic turbulence prediction, there are also many pairs of identical contributions from different terms in Table 3. Compared with the low wave number cases, the inertial region pressure spectra need more terms as the main contributions.

**Table 3. Main contributions to Run 1 t-t interaction pressure fluctuation spectra at $k_1 = 10\,m^{-1}$ for anisotropic turbulence calculation at 1 meter and calculation with isotropic turbulence input.**[*]

| $k_1 = 10\,m^{-1}$ | $F_{11}F_{11}$ ($\times 10^{-3}$ $Pa^2 m$) | $F_{33}F_{33}$ ($\times 10^{-3}$ $Pa^2 m$) | $F_{33}F_{11}$ ($\times 10^{-3}$ $Pa^2 m$) | $F_{11}F_{33}$ ($\times 10^{-3}$ $Pa^2 m$) | $F_{13}F_{33}$ ($\times 10^{-3}$ $Pa^2 m$) | $F_{33}F_{13}$ ($\times 10^{-3}$ $Pa^2 m$) | $F_{13}F_{11}$ ($\times 10^{-3}$ $Pa^2 m$) | $F_{11}F_{13}$ ($\times 10^{-3}$ $Pa^2 m$) | $F_{13}F_{13}$ ($\times 10^{-3}$ $Pa^2 m$) |
|---|---|---|---|---|---|---|---|---|---|
| Anisotropic @1 m | 3.227 | 1.847 | 1.959 | 1.960 | -1.502 | -0.747 | -1.726 | -1.350 | 0.114 |
| Isotropic input | 3.227 | 1.847 | 1.959 | 1.960 | -1.502 | -0.747 | -1.726 | -1.350 | 0.114 |
| Isotropic (George) | - | - | - | - | - | - | - | - | - |
| $k_1 = 10\,m^{-1}$ | $F_{23}F_{33}$ ($\times 10^{-3}$ $Pa^2 m$) | $F_{21}F_{11}$ ($\times 10^{-3}$ $Pa^2 m$) | $F_{11}F_{21}$ ($\times 10^{-3}$ $Pa^2 m$) | $F_{23}F_{11}$ ($\times 10^{-3}$ $Pa^2 m$) | $F_{21}F_{33}$ ($\times 10^{-3}$ $Pa^2 m$) | $F_{33}F_{21}$ ($\times 10^{-3}$ $Pa^2 m$) | $F_{11}F_{23}$ ($\times 10^{-3}$ $Pa^2 m$) | $F_{33}F_{23}$ ($\times 10^{-3}$ $Pa^2 m$) | |
| Anisotropic @1 m | -1.191 | -1.154 | -0.858 | -0.363 | -0.338 | -0.243 | -0.232 | -0.223 | |
| Isotropic input | -1.840 | -1.726 | -1.350 | -0.602 | -0.504 | -0.380 | -0.330 | -0.291 | |
| Isotropic (George) | - | - | - | - | - | - | - | - | |
| $k_1 = 10\,m^{-1}$ | $F_{11}F_{22}$ ($\times 10^{-3}$ $Pa^2 m$) | $F_{22}F_{11}$ ($\times 10^{-3}$ $Pa^2 m$) | $F_{33}F_{22}$ ($\times 10^{-3}$ $Pa^2 m$) | $F_{22}F_{33}$ ($\times 10^{-3}$ $Pa^2 m$) | $F_{13}F_{22}$ ($\times 10^{-3}$ $Pa^2 m$) | $F_{22}F_{13}$ ($\times 10^{-3}$ $Pa^2 m$) | $F_{23}F_{23}$ ($\times 10^{-3}$ $Pa^2 m$) | $F_{21}F_{21}$ ($\times 10^{-3}$ $Pa^2 m$) | |
| Anisotropic @1 m | 1.228 | 1.312 | 0.631 | 0.922 | -0.234 | -0.112 | 1.007 | 0.037 | |
| Isotropic input | 1.957 | 1.959 | 1.074 | 1.074 | -0.504 | -0.380 | 1.417 | 0.114 | |
| Isotropic (George) | - | - | - | - | - | - | - | - | |
| $k_1 = 10\,m^{-1}$ | $F_{22}F_{22}$ ($\times 10^{-3}$ $Pa^2 m$) | $F_{23}F_{22}$ ($\times 10^{-3}$ $Pa^2 m$) | $F_{21}F_{22}$ ($\times 10^{-3}$ $Pa^2 m$) | $F_{22}F_{21}$ ($\times 10^{-3}$ $Pa^2 m$) | $F_{22}F_{23}$ ($\times 10^{-3}$ $Pa^2 m$) | Total in Fig. 1(c) ($\times 10^{-3}$ $Pa^2 m$) | | | |
| Anisotropic @1 m | 0.910 | -0.903 | -0.656 | -0.320 | -0.108 | 3.16 | | | |
| Isotropic input | 1.847 | -1.840 | -1.502 | -0.747 | -0.291 | 1.33 | | | |
| Isotropic (George) | - | - | - | - | - | 1.33 | | | |

[*]Note: Contributions with absolute values below $10^{-4}\,Pa^2\,m$ are not listed.

It was found [4,6] that for homogeneous isotropic turbulence, the three dimensional turbulence-turbulence interaction pressure spectrum $F_{p,p}^t(\vec{k})$ and one of the fourth-order velocity spectra are equivalent:

$$\frac{1}{\rho^2} F_{p,p}^t(\vec{k}) = F_{LL,LL}(k), \tag{17}$$

where $F_{LL,LL}$ is a function of $k\ (=|\vec{k}|)$, and the subscript $L$ indicates that the Fourier velocity component $u_L$ is aligned with the wave number vector $\vec{k}$. One may henceforth conjecture that the one dimensional turbulence-turbulence interaction pressure spectrum $F_{ppt}^1(k_1)$ should be identical with $\rho^2 F_{11,11}(k_1)$, and the one dimensional turbulence-turbulence interaction pressure spectrum should be equivalent with the contribution from the $F_{11}F_{11}$ term only. The conjecture above is not correct. Table 3 shows the evidence. From Table 3, at $k_1 = 10\,m^{-1}$, the one dimensional turbulence-turbulence interaction pressure spectrum for isotropic turbulence has a value of $1.33\times10^{-3}\,Pa^2\,m$, and the contribution from the $F_{11}F_{11}$ term has a value of $3.227\times10^{-3}\,Pa^2\,m$. They are not equivalent. Caution is needed with the notation when trying to derive $F_{ppt}^1(k_1)$ from Eq. 17. Appendix E shows the derivation process for the one dimensional turbulence-turbulence interaction pressure spectrum from Eq. 17. Eq. E-7 is the derived form of the one dimensional turbulence-turbulence interaction pressure spectrum from Eq. 17. Eq. E-7 yields identical value with the value in Table 3 calculated using Eq. D-3 (a value of $1.33\times10^{-3}\,Pa^2\,m$ at $k_1 = 10\,m^{-1}$). Hence George et al.'s finding (Eq. 17) and the inconsistency in Table 3 between contribution from the $F_{11}F_{11}$ term and the one dimensional pressure spectral value for isotropic turbulence are not contradictory. Appendix E interprets the relations between Eq. E-7 and contribution from the $F_{11}F_{11}$ term in Appendix D for isotropic input calculation (Eq. E-8) based on their expressions. We can see despite they are different forms which should yield different values, they are reasonably related.

## IV. CONCLUSION

The method for the incorporation of anisotropic turbulence into the calculation of the turbulence-turbulence interaction pressure above the ground has been developed for wind noise pressure spectral predictions under homogeneous anisotropic turbulence conditions. The turbulence anisotropy retains the constant spectral level of the t-t interaction pressure spectrum at low wave number, but alters the spectral slope in inertial region in the logarithmic coordinate to about -5/3. The turbulence anisotropy's influence to the turbulence-turbulence interaction pressure spectrum in the source region is relatively small, but it yields a higher level in the inertial region. Since the t-t contribution is relatively more important than the t-s contribution in the inertial region to the total pressure spectrum as wavenumber increases, turbulence anisotropy is thought to have non-negligible influence on the total pressure spectral prediction in the inertial region.

The turbulence-turbulence interaction pressure spectrum by incorporating turbulence anisotropy is not sensitive to height. The t-t interaction pressure spectrum using the method we developed but with homogeneous isotropic input matches the t-t interaction pressure spectrum for homogeneous isotropic turbulence using George et al.'s method [4] except at the high wave number end of the data range. Different contributions to the turbulence-turbulence interaction pressure spectra are evaluated for the anisotropic and isotropic turbulence models at 1 meter elevation. Dominant contributions are identified for different wave number regions. The slope on logarithmic axes becomes smaller for the anisotropic t-t interaction pressure spectrum in the inertial region compared with isotropic turbulence because it has less negative value pressure contributions which fall off more rapidly. Finally, the one dimensional t-t interaction pressure spectrum is derived from the relation found by George et al. for the fourth-order velocity spectra under homogeneous isotropic turbulence conditions. The derived form gives pressure spectral value identical with the isotropic turbulence pressure spectral calculation, which validates George et al.'s argument on the relation of pressure spectrum and one of the fourth-order velocity spectra, and clarifies the difference between the $F_{11}F_{11}$ term contribution and the isotropic turbulence pressure calculation.

Based on the results we obtained and the high degree of consistency between calculations using George et al's method and ours with the isotropic turbulence input, the reliability and justification of the methods we developed based on Kraichnan's mirror flow model for pressure spectral predictions under anisotropic turbulence conditions are implied. This theoretical approach to calculating the contribution of the turbulence-turbulence interaction to wind noise pressure spectra has demonstrated the consequences of anisotropy with respect to the homogeneous isotropic case. These results offer insight regarding the role of turbulence anisotropy in the turbulence-turbulence interaction pressure fluctuation spectra in the atmospheric surface layer.

**References**

1. Yu, J., Raspet, R., Webster, J., Abbott, J.P.: Wind noise measured at the ground surface. J. Acoust. Soc. Am. **129**(2), 622-632 (2011)

APPENDIX A.   Conversion Forms for $F(\widetilde{R}_{\alpha\beta} \cdot \widetilde{R}_{\mu\nu})$

For $(\widetilde{R}_{\alpha\beta} \cdot \widetilde{R}_{\mu\nu}) = (\widetilde{R}_{11} \cdot \widetilde{R}_{23}), (\widetilde{R}_{33} \cdot \widetilde{R}_{23}), (\widetilde{R}_{11} \cdot \widetilde{R}_{21}), (\widetilde{R}_{33} \cdot \widetilde{R}_{21}), (\widetilde{R}_{13} \cdot \widetilde{R}_{21}), (\widetilde{R}_{13} \cdot \widetilde{R}_{23})$,

$$F(\widetilde{R}_{\alpha\beta} \cdot \widetilde{R}_{\mu\nu})$$
$$= \frac{1}{(2\pi)^{3/2}} \int \left( \frac{1}{\sqrt{2\pi}} \int R_{\alpha\beta}(\vec{k} - \vec{k}') e^{i(k_2 - k_2')(x_2' - x_2)} d(k_2 - k_2') \right) \left( \frac{1}{\sqrt{2}} \frac{1}{\sqrt{2\pi}} \int R_{\mu\nu}(\vec{k}')(e^{ik_2'(x_2' - x_2)} - e^{ik_2'(-x_2' - x_2)}) dk_2' \right) d^2\vec{\kappa}'$$
$$= \frac{1}{8\pi^{5/2}} [\int e^{ik_2(x_2' - x_2)} \left( \int R_{\alpha\beta}(\vec{k} - \vec{k}') R_{\mu\nu}(\vec{k}') d^3\vec{k}' \right) dk_2 - \int e^{ik_2(x_2' - x_2)} \left( \int e^{-2ik_2'x_2'} R_{\alpha\beta}(\vec{k} - \vec{k}') R_{\mu\nu}(\vec{k}') d^3\vec{k}' \right) dk_2]$$

(A-1)

For $(\widetilde{R}_{\alpha\beta} \cdot \widetilde{R}_{\mu\nu}) = (\widetilde{R}_{23} \cdot \widetilde{R}_{11}), (\widetilde{R}_{23} \cdot \widetilde{R}_{33}), (\widetilde{R}_{21} \cdot \widetilde{R}_{11}), (\widetilde{R}_{21} \cdot \widetilde{R}_{33}), (\widetilde{R}_{21} \cdot \widetilde{R}_{13}), (\widetilde{R}_{23} \cdot \widetilde{R}_{13})$,

$$F(\widetilde{R}_{\alpha\beta} \cdot \widetilde{R}_{\mu\nu})$$
$$= \frac{1}{8\pi^{5/2}} [\int e^{ik_2(x_2' - x_2)} \left( \int R_{\alpha\beta}(\vec{k} - \vec{k}') R_{\mu\nu}(\vec{k}') d^3\vec{k}' \right) dk_2 - \int e^{-ik_2(x_2' + x_2)} \left( \int e^{2ik_2'x_2'} R_{\alpha\beta}(\vec{k} - \vec{k}') R_{\mu\nu}(\vec{k}') d^3\vec{k}' \right) dk_2]$$

(A-2)

For $(\widetilde{R}_{\alpha\beta} \cdot \widetilde{R}_{\mu\nu}) = (\widetilde{R}_{11} \cdot \widetilde{R}_{32}), (\widetilde{R}_{33} \cdot \widetilde{R}_{32}), (\widetilde{R}_{11} \cdot \widetilde{R}_{12}), (\widetilde{R}_{33} \cdot \widetilde{R}_{12}), (\widetilde{R}_{13} \cdot \widetilde{R}_{12}), (\widetilde{R}_{13} \cdot \widetilde{R}_{32})$,

$$F(\widetilde{R}_{\alpha\beta} \cdot \widetilde{R}_{\mu\nu})$$
$$= \frac{1}{(2\pi)^{3/2}} \int \left( \frac{1}{\sqrt{2\pi}} \int R_{\alpha\beta}(\vec{k} - \vec{k}') e^{i(k_2 - k_2')(x_2' - x_2)} d(k_2 - k_2') \right) \left( \frac{1}{\sqrt{2}} \frac{1}{\sqrt{2\pi}} \int R_{\mu\nu}(\vec{k}')(e^{ik_2'(x_2' - x_2)} - e^{ik_2'(x_2' + x_2)}) dk_2' \right) d^2\vec{\kappa}'$$
$$= \frac{1}{8\pi^{5/2}} [\int e^{ik_2(x_2' - x_2)} \left( \int R_{\alpha\beta}(\vec{k} - \vec{k}') R_{\mu\nu}(\vec{k}') d^3\vec{k}' \right) dk_2 - \int e^{ik_2(x_2' - x_2)} \left( \int e^{2ik_2'x_2} R_{\alpha\beta}(\vec{k} - \vec{k}') R_{\mu\nu}(\vec{k}') d^3\vec{k}' \right) dk_2]$$

(A-3)

For $(\widetilde{R}_{\alpha\beta} \cdot \widetilde{R}_{\mu\nu}) = (\widetilde{R}_{32} \cdot \widetilde{R}_{11}), (\widetilde{R}_{32} \cdot \widetilde{R}_{33}), (\widetilde{R}_{12} \cdot \widetilde{R}_{11}), (\widetilde{R}_{12} \cdot \widetilde{R}_{33}), (\widetilde{R}_{12} \cdot \widetilde{R}_{13}), (\widetilde{R}_{32} \cdot \widetilde{R}_{13})$,

$$F(\widetilde{R}_{\alpha\beta} \cdot \widetilde{R}_{\mu\nu})$$
$$= \frac{1}{8\pi^{5/2}}[\int e^{ik_2(x_2'-x_2)} \left(\int R_{\alpha\beta}(\vec{k}-\vec{k}')R_{\mu\nu}(\vec{k}')d^3\vec{k}'\right)dk_2 - \int e^{ik_2(x_2'+x_2)} \left(\int e^{-2ik_2'x_2} R_{\alpha\beta}(\vec{k}-\vec{k}')R_{\mu\nu}(\vec{k}')d^3\vec{k}'\right)dk_2]$$

(A-4)

For $(\widetilde{R}_{\alpha\beta} \cdot \widetilde{R}_{\mu\nu}) = (\widetilde{R}_{11} \cdot \widetilde{R}_{22}), (\widetilde{R}_{33} \cdot \widetilde{R}_{22}), (\widetilde{R}_{13} \cdot \widetilde{R}_{22})$,

$$F(\widetilde{R}_{\alpha\beta} \cdot \widetilde{R}_{\mu\nu})$$
$$= \frac{1}{(2\pi)^{3/2}} \int \left(\frac{1}{\sqrt{2\pi}} \int R_{\alpha\beta}(\vec{k}-\vec{k}')e^{i(k_2-k_2')(x_2'-x_2)}d(k_2-k_2')\right)\left(\frac{1}{2}\frac{1}{\sqrt{2\pi}} \int R_{\mu\nu}(\vec{k}')(e^{ik_2'(x_2'-x_2)} + e^{-ik_2'(x_2'-x_2)} - e^{ik_2'(x_2'+x_2)} - e^{-ik_2'(x_2'+x_2)})dk_2'\right)d^2\vec{\kappa}'$$
$$= \frac{1}{8\sqrt{2}\pi^{5/2}}[\int e^{ik_2(x_2'-x_2)} \left(\int R_{\alpha\beta}(\vec{k}-\vec{k}')R_{\mu\nu}(\vec{k}')d^3\vec{k}'\right)dk_2 + \int e^{ik_2(x_2'-x_2)} \left(\int e^{-2ik_2'(x_2'-x_2)} R_{\alpha\beta}(\vec{k}-\vec{k}')R_{\mu\nu}(\vec{k}')d^3\vec{k}'\right)dk_2$$
$$- \int e^{ik_2(x_2'-x_2)} \left(\int e^{2ik_2'x_2} R_{\alpha\beta}(\vec{k}-\vec{k}')R_{\mu\nu}(\vec{k}')d^3\vec{k}'\right)dk_2 - \int e^{ik_2(x_2'-x_2)} \left(\int e^{-2ik_2'x_2} R_{\alpha\beta}(\vec{k}-\vec{k}')R_{\mu\nu}(\vec{k}')d^3\vec{k}'\right)dk_2]$$

(A-5)

For $(\widetilde{R}_{\alpha\beta} \cdot \widetilde{R}_{\mu\nu}) = (\widetilde{R}_{22} \cdot \widetilde{R}_{11}), (\widetilde{R}_{22} \cdot \widetilde{R}_{33}), (\widetilde{R}_{22} \cdot \widetilde{R}_{13})$,

$$F(\widetilde{R}_{\alpha\beta} \cdot \widetilde{R}_{\mu\nu})$$
$$= \frac{1}{8\sqrt{2}\pi^{5/2}}[\int e^{ik_2(x_2'-x_2)} \left(\int R_{\alpha\beta}(\vec{k}-\vec{k}')R_{\mu\nu}(\vec{k}')d^3\vec{k}'\right)dk_2 + \int e^{-ik_2(x_2'-x_2)} \left(\int e^{2ik_2'(x_2'-x_2)} R_{\alpha\beta}(\vec{k}-\vec{k}')R_{\mu\nu}(\vec{k}')d^3\vec{k}'\right)dk_2$$
$$- \int e^{ik_2(x_2'+x_2)} \left(\int e^{-2ik_2'x_2} R_{\alpha\beta}(\vec{k}-\vec{k}')R_{\mu\nu}(\vec{k}')d^3\vec{k}'\right)dk_2 - \int e^{-ik_2(x_2'+x_2)} \left(\int e^{2ik_2'x_2} R_{\alpha\beta}(\vec{k}-\vec{k}')R_{\mu\nu}(\vec{k}')d^3\vec{k}'\right)dk_2]$$

(A-6)

For $(\widetilde{R}_{\alpha\beta} \cdot \widetilde{R}_{\mu\nu}) = (\widetilde{R}_{21} \cdot \widetilde{R}_{21}), (\widetilde{R}_{23} \cdot \widetilde{R}_{23}), (\widetilde{R}_{21} \cdot \widetilde{R}_{23}), (\widetilde{R}_{23} \cdot \widetilde{R}_{21})$,

$$F(\widetilde{R}_{\alpha\beta} \cdot \widetilde{R}_{\mu\nu})$$
$$= \frac{1}{(2\pi)^{3/2}} \int \left(\frac{1}{\sqrt{2}}\frac{1}{\sqrt{2\pi}} \int R_{\alpha\beta}(\vec{k}-\vec{k}')(e^{i(k_2-k_2')(x_2'-x_2)} - e^{i(k_2-k_2')(-x_2'-x_2)})d(k_2-k_2')\right)\left(\frac{1}{\sqrt{2}}\frac{1}{\sqrt{2\pi}} \int R_{\mu\nu}(\vec{k}')(e^{ik_2'(x_2'-x_2)} - e^{ik_2'(-x_2'-x_2)})dk_2'\right)d^2\vec{\kappa}'$$
$$= \frac{1}{8\sqrt{2}\pi^{5/2}}[\int e^{ik_2(x_2'-x_2)} \left(\int R_{\alpha\beta}(\vec{k}-\vec{k}')R_{\mu\nu}(\vec{k}')d^3\vec{k}'\right)dk_2 - \int e^{ik_2(x_2'-x_2)} \left(\int e^{-2ik_2'x_2} R_{\alpha\beta}(\vec{k}-\vec{k}')R_{\mu\nu}(\vec{k}')d^3\vec{k}'\right)dk_2$$
$$- \int e^{-ik_2(x_2'+x_2)} \left(\int e^{2ik_2'x_2} R_{\alpha\beta}(\vec{k}-\vec{k}')R_{\mu\nu}(\vec{k}')d^3\vec{k}'\right)dk_2 + \int e^{-ik_2(x_2'+x_2)} \left(\int R_{\alpha\beta}(\vec{k}-\vec{k}')R_{\mu\nu}(\vec{k}')d^3\vec{k}'\right)dk_2]$$

(A-7)

For $(\widetilde{R}_{\alpha\beta} \cdot \widetilde{R}_{\mu\nu}) = (\widetilde{R}_{12} \cdot \widetilde{R}_{12}), (\widetilde{R}_{32} \cdot \widetilde{R}_{32}), (\widetilde{R}_{12} \cdot \widetilde{R}_{32}), (\widetilde{R}_{32} \cdot \widetilde{R}_{12})$,

$$F(\widetilde{R}_{\alpha\beta} \cdot \widetilde{R}_{\mu\nu})$$
$$= \frac{1}{8\sqrt{2}\pi^{5/2}}[\int e^{ik_2(x_2'-x_2)} \left(\int R_{\alpha\beta}(\vec{k}-\vec{k}')R_{\mu\nu}(\vec{k}')d^3\vec{k}'\right)dk_2 - \int e^{ik_2(x_2'-x_2)} \left(\int e^{2ik_2'x_2} R_{\alpha\beta}(\vec{k}-\vec{k}')R_{\mu\nu}(\vec{k}')d^3\vec{k}'\right)dk_2$$
$$- \int e^{ik_2(x_2'+x_2)} \left(\int e^{-2ik_2'x_2} R_{\alpha\beta}(\vec{k}-\vec{k}')R_{\mu\nu}(\vec{k}')d^3\vec{k}'\right)dk_2 + \int e^{ik_2(x_2'+x_2)} \left(\int R_{\alpha\beta}(\vec{k}-\vec{k}')R_{\mu\nu}(\vec{k}')d^3\vec{k}'\right)dk_2]$$

(A-8)

For $(\widetilde{R}_{\alpha\beta} \cdot \widetilde{R}_{\mu\nu}) = (\widetilde{R}_{12} \cdot \widetilde{R}_{21}), (\widetilde{R}_{32} \cdot \widetilde{R}_{23}), (\widetilde{R}_{12} \cdot \widetilde{R}_{23}), (\widetilde{R}_{32} \cdot \widetilde{R}_{21})$,

$$F(\widetilde{R}_{\alpha\beta} \cdot \widetilde{R}_{\mu\nu})$$
$$= \frac{1}{8\sqrt{2}\pi^{5/2}}[\int e^{ik_2(x_2'-x_2)} \left(\int R_{\alpha\beta}(\vec{k}-\vec{k}')R_{\mu\nu}(\vec{k}')d^3\vec{k}'\right)dk_2 - \int e^{ik_2(x_2'-x_2)} \left(\int e^{-2ik_2'x_2} R_{\alpha\beta}(\vec{k}-\vec{k}')R_{\mu\nu}(\vec{k}')d^3\vec{k}'\right)dk_2$$
$$- \int e^{ik_2(x_2'+x_2)} \left(\int e^{-2ik_2'x_2} R_{\alpha\beta}(\vec{k}-\vec{k}')R_{\mu\nu}(\vec{k}')d^3\vec{k}'\right)dk_2 + \int e^{ik_2(x_2'+x_2)} \left(\int e^{-2ik_2'(x_2'+x_2)} R_{\alpha\beta}(\vec{k}-\vec{k}')R_{\mu\nu}(\vec{k}')d^3\vec{k}'\right)dk_2]$$

(A-9)

For $(\widetilde{R}_{\alpha\beta} \cdot \widetilde{R}_{\mu\nu}) = (\widetilde{R}_{21} \cdot \widetilde{R}_{12}), (\widetilde{R}_{23} \cdot \widetilde{R}_{32}), (\widetilde{R}_{21} \cdot \widetilde{R}_{32}), (\widetilde{R}_{23} \cdot \widetilde{R}_{12})$,

$$F(\widetilde{R}_{\alpha\beta} \cdot \widetilde{R}_{\mu\nu})$$

$$= \frac{1}{8\sqrt{2}\pi^{5/2}} [\int e^{ik_2(x_2'-x_2)} \left(\int R_{\alpha\beta}(\vec{k}-\vec{k}')R_{\mu\nu}(\vec{k}')d^3\vec{k}'\right) dk_2 - \int e^{ik_2(x_2'-x_2)} \left(\int e^{2ik_2'x_2} R_{\alpha\beta}(\vec{k}-\vec{k}')R_{\mu\nu}(\vec{k}')d^3\vec{k}'\right) dk_2$$

$$- \int e^{-ik_2(x_2'+x_2)} \left(\int e^{2ik_2'x_2} R_{\alpha\beta}(\vec{k}-\vec{k}')R_{\mu\nu}(\vec{k}')d^3\vec{k}'\right) dk_2 + \int e^{-ik_2(x_2'+x_2)} \left(\int e^{2ik_2'(x_2'+x_2)} R_{\alpha\beta}(\vec{k}-\vec{k}')R_{\mu\nu}(\vec{k}')d^3\vec{k}'\right) dk_2]$$

(A-10)

For $(\widetilde{R}_{\alpha\beta} \cdot \widetilde{R}_{\mu\nu}) = (\widetilde{R}_{22} \cdot \widetilde{R}_{21}), (\widetilde{R}_{22} \cdot \widetilde{R}_{23})$,

$$F(\widetilde{R}_{\alpha\beta} \cdot \widetilde{R}_{\mu\nu})$$

$$= \frac{1}{16\pi^{5/2}} [\int e^{ik_2(x_2'-x_2)} \left(\int R_{\alpha\beta}(\vec{k}-\vec{k}')R_{\mu\nu}(\vec{k}')d^3\vec{k}'\right) dk_2 - \int e^{ik_2(x_2'-x_2)} \left(\int e^{-2ik_2'x_2} R_{\alpha\beta}(\vec{k}-\vec{k}')R_{\mu\nu}(\vec{k}')d^3\vec{k}'\right) dk_2$$

$$+ \int e^{-ik_2(x_2'-x_2)} \left(\int e^{2ik_2'(x_2'-x_2)} R_{\alpha\beta}(\vec{k}-\vec{k}')R_{\mu\nu}(\vec{k}')d^3\vec{k}'\right) dk_2 - \int e^{-ik_2(x_2'-x_2)} \left(\int e^{-2ik_2'x_2} R_{\alpha\beta}(\vec{k}-\vec{k}')R_{\mu\nu}(\vec{k}')d^3\vec{k}'\right) dk_2$$

$$- \int e^{ik_2(x_2'+x_2)} \left(\int e^{-2ik_2'x_2} R_{\alpha\beta}(\vec{k}-\vec{k}')R_{\mu\nu}(\vec{k}')d^3\vec{k}'\right) dk_2 + \int e^{ik_2(x_2'+x_2)} \left(\int e^{-2ik_2'(x_2'+x_2)} R_{\alpha\beta}(\vec{k}-\vec{k}')R_{\mu\nu}(\vec{k}')d^3\vec{k}'\right) dk_2$$

$$- \int e^{-ik_2(x_2'+x_2)} \left(\int e^{2ik_2'x_2} R_{\alpha\beta}(\vec{k}-\vec{k}')R_{\mu\nu}(\vec{k}')d^3\vec{k}'\right) dk_2 + \int e^{-ik_2(x_2'+x_2)} \left(\int R_{\alpha\beta}(\vec{k}-\vec{k}')R_{\mu\nu}(\vec{k}')d^3\vec{k}'\right) dk_2]$$

(A-11)

For $(\widetilde{R}_{\alpha\beta} \cdot \widetilde{R}_{\mu\nu}) = (\widetilde{R}_{22} \cdot \widetilde{R}_{12}), (\widetilde{R}_{22} \cdot \widetilde{R}_{32})$,

$$F(\widetilde{R}_{\alpha\beta} \cdot \widetilde{R}_{\mu\nu})$$

$$= \frac{1}{16\pi^{5/2}} [\int e^{ik_2(x_2'-x_2)} \left(\int R_{\alpha\beta}(\vec{k}-\vec{k}')R_{\mu\nu}(\vec{k}')d^3\vec{k}'\right) dk_2 - \int e^{ik_2(x_2'-x_2)} \left(\int e^{2ik_2'x_2} R_{\alpha\beta}(\vec{k}-\vec{k}')R_{\mu\nu}(\vec{k}')d^3\vec{k}'\right) dk_2$$

$$+ \int e^{-ik_2(x_2'-x_2)} \left(\int e^{2ik_2'(x_2'-x_2)} R_{\alpha\beta}(\vec{k}-\vec{k}')R_{\mu\nu}(\vec{k}')d^3\vec{k}'\right) dk_2 - \int e^{-ik_2(x_2'-x_2)} \left(\int e^{2ik_2'x_2} R_{\alpha\beta}(\vec{k}-\vec{k}')R_{\mu\nu}(\vec{k}')d^3\vec{k}'\right) dk_2$$

$$- \int e^{ik_2(x_2'+x_2)} \left(\int e^{-2ik_2'x_2} R_{\alpha\beta}(\vec{k}-\vec{k}')R_{\mu\nu}(\vec{k}')d^3\vec{k}'\right) dk_2 + \int e^{ik_2(x_2'+x_2)} \left(\int R_{\alpha\beta}(\vec{k}-\vec{k}')R_{\mu\nu}(\vec{k}')d^3\vec{k}'\right) dk_2$$

$$- \int e^{-ik_2(x_2'+x_2)} \left(\int e^{2ik_2'x_2} R_{\alpha\beta}(\vec{k}-\vec{k}')R_{\mu\nu}(\vec{k}')d^3\vec{k}'\right) dk_2 + \int e^{-ik_2(x_2'+x_2)} \left(\int e^{2ik_2'(x_2'+x_2)} R_{\alpha\beta}(\vec{k}-\vec{k}')R_{\mu\nu}(\vec{k}')d^3\vec{k}'\right) dk_2]$$

(A-12)

For $(\widetilde{R}_{\alpha\beta} \cdot \widetilde{R}_{\mu\nu}) = (\widetilde{R}_{21} \cdot \widetilde{R}_{22}), (\widetilde{R}_{23} \cdot \widetilde{R}_{22})$,

$$F(\widetilde{R}_{\alpha\beta} \cdot \widetilde{R}_{\mu\nu})$$

$$= \frac{1}{16\pi^{5/2}} [\int e^{ik_2(x_2'-x_2)} \left(\int R_{\alpha\beta}(\vec{k}-\vec{k}')R_{\mu\nu}(\vec{k}')d^3\vec{k}'\right) dk_2 + \int e^{ik_2(x_2'-x_2)} \left(\int e^{-2ik_2'(x_2'-x_2)} R_{\alpha\beta}(\vec{k}-\vec{k}')R_{\mu\nu}(\vec{k}')d^3\vec{k}'\right) dk_2$$

$$- \int e^{ik_2(x_2'-x_2)} \left(\int e^{2ik_2'x_2} R_{\alpha\beta}(\vec{k}-\vec{k}')R_{\mu\nu}(\vec{k}')d^3\vec{k}'\right) dk_2 - \int e^{-ik_2(x_2'-x_2)} \left(\int e^{-2ik_2'x_2} R_{\alpha\beta}(\vec{k}-\vec{k}')R_{\mu\nu}(\vec{k}')d^3\vec{k}'\right) dk_2$$

$$- \int e^{-ik_2(x_2'+x_2)} \left(\int e^{2ik_2'x_2} R_{\alpha\beta}(\vec{k}-\vec{k}')R_{\mu\nu}(\vec{k}')d^3\vec{k}'\right) dk_2 - \int e^{-ik_2(x_2'+x_2)} \left(\int e^{2ik_2'x_2} R_{\alpha\beta}(\vec{k}-\vec{k}')R_{\mu\nu}(\vec{k}')d^3\vec{k}'\right) dk_2$$

$$+ \int e^{-ik_2(x_2'+x_2)} \left(\int e^{2ik_2'(x_2'+x_2)} R_{\alpha\beta}(\vec{k}-\vec{k}')R_{\mu\nu}(\vec{k}')d^3\vec{k}'\right) dk_2 + \int e^{-ik_2(x_2'+x_2)} \left(\int R_{\alpha\beta}(\vec{k}-\vec{k}')R_{\mu\nu}(\vec{k}')d^3\vec{k}'\right) dk_2]$$

(A-13)

For $(\widetilde{R}_{\alpha\beta} \cdot \widetilde{R}_{\mu\nu}) = (\widetilde{R}_{12} \cdot \widetilde{R}_{22}), (\widetilde{R}_{32} \cdot \widetilde{R}_{22})$,

$$F(\widetilde{R}_{\alpha\beta} \cdot \widetilde{R}_{\mu\nu})$$

$$= \frac{1}{16\pi^{5/2}} [\int e^{ik_2(x_2'-x_2)} \left(\int R_{\alpha\beta}(\vec{k}-\vec{k}')R_{\mu\nu}(\vec{k}')d^3\vec{k}'\right) dk_2 + \int e^{ik_2(x_2'-x_2)} \left(\int e^{-2ik_2'(x_2'-x_2)} R_{\alpha\beta}(\vec{k}-\vec{k}')R_{\mu\nu}(\vec{k}')d^3\vec{k}'\right) dk_2$$

$$- \int e^{ik_2(x_2'-x_2)} \left(\int e^{2ik_2'x_2} R_{\alpha\beta}(\vec{k}-\vec{k}')R_{\mu\nu}(\vec{k}')d^3\vec{k}'\right) dk_2 - \int e^{-ik_2(x_2'-x_2)} \left(\int e^{-2ik_2'x_2} R_{\alpha\beta}(\vec{k}-\vec{k}')R_{\mu\nu}(\vec{k}')d^3\vec{k}'\right) dk_2$$

$$- \int e^{ik_2(x_2'+x_2)} \left(\int e^{-2ik_2'x_2} R_{\alpha\beta}(\vec{k}-\vec{k}')R_{\mu\nu}(\vec{k}')d^3\vec{k}'\right) dk_2 - \int e^{ik_2(x_2'+x_2)} \left(\int e^{-2ik_2'x_2} R_{\alpha\beta}(\vec{k}-\vec{k}')R_{\mu\nu}(\vec{k}')d^3\vec{k}'\right) dk_2$$

$$+ \int e^{ik_2(x_2'+x_2)} \left(\int R_{\alpha\beta}(\vec{k}-\vec{k}')R_{\mu\nu}(\vec{k}')d^3\vec{k}'\right) dk_2 + \int e^{ik_2(x_2'+x_2)} \left(\int e^{-2ik_2'(x_2'+x_2)} R_{\alpha\beta}(\vec{k}-\vec{k}')R_{\mu\nu}(\vec{k}')d^3\vec{k}'\right) dk_2]$$

(A-14)

For $(\widetilde{R}_{\alpha\beta} \cdot \widetilde{R}_{\mu\nu}) = (\widetilde{R}_{22} \cdot \widetilde{R}_{22})$,

$F(\widetilde{R}_{\alpha\beta} \cdot \widetilde{R}_{\mu\nu})$

$= \dfrac{1}{16\sqrt{2}\pi^{5/2}} [\int e^{ik_2(x_2'-x_2)} \left(\int R_{\alpha\beta}(\vec{k}-\vec{k}')R_{\mu\nu}(\vec{k}')d^3\vec{k}'\right)dk_2 + \int e^{ik_2(x_2'-x_2)} \left(\int e^{-2ik_2'(x_2'-x_2)} R_{\alpha\beta}(\vec{k}-\vec{k}')R_{\mu\nu}(\vec{k}')d^3\vec{k}'\right)dk_2$

$- \int e^{ik_2(x_2'-x_2)} \left(\int e^{2ik_2'x_2} R_{\alpha\beta}(\vec{k}-\vec{k}')R_{\mu\nu}(\vec{k}')d^3\vec{k}'\right)dk_2 - \int e^{ik_2(x_2'-x_2)} \left(\int e^{-2ik_2'x_2} R_{\alpha\beta}(\vec{k}-\vec{k}')R_{\mu\nu}(\vec{k}')d^3\vec{k}'\right)dk_2$

$+ \int e^{-ik_2(x_2'-x_2)} \left(\int e^{2ik_2'(x_2'-x_2)} R_{\alpha\beta}(\vec{k}-\vec{k}')R_{\mu\nu}(\vec{k}')d^3\vec{k}'\right)dk_2 + \int e^{-ik_2(x_2'-x_2)} \left(\int R_{\alpha\beta}(\vec{k}-\vec{k}')R_{\mu\nu}(\vec{k}')d^3\vec{k}'\right)dk_2$

$- \int e^{-ik_2(x_2'-x_2)} \left(\int e^{2ik_2'x_2} R_{\alpha\beta}(\vec{k}-\vec{k}')R_{\mu\nu}(\vec{k}')d^3\vec{k}'\right)dk_2 - \int e^{-ik_2(x_2'-x_2)} \left(\int e^{-2ik_2'x_2} R_{\alpha\beta}(\vec{k}-\vec{k}')R_{\mu\nu}(\vec{k}')d^3\vec{k}'\right)dk_2$

$- \int e^{ik_2(x_2'+x_2)} \left(\int e^{-2ik_2'x_2} R_{\alpha\beta}(\vec{k}-\vec{k}')R_{\mu\nu}(\vec{k}')d^3\vec{k}'\right)dk_2 - \int e^{ik_2(x_2'+x_2)} \left(\int e^{-2ik_2'x_2} R_{\alpha\beta}(\vec{k}-\vec{k}')R_{\mu\nu}(\vec{k}')d^3\vec{k}'\right)dk_2$

$+ \int e^{ik_2(x_2'+x_2)} \left(\int R_{\alpha\beta}(\vec{k}-\vec{k}')R_{\mu\nu}(\vec{k}')d^3\vec{k}'\right)dk_2 + \int e^{ik_2(x_2'+x_2)} \left(\int e^{-2ik_2'(x_2'+x_2)} R_{\alpha\beta}(\vec{k}-\vec{k}')R_{\mu\nu}(\vec{k}')d^3\vec{k}'\right)dk_2$

$- \int e^{-ik_2(x_2'+x_2)} \left(\int e^{2ik_2'x_2} R_{\alpha\beta}(\vec{k}-\vec{k}')R_{\mu\nu}(\vec{k}')d^3\vec{k}'\right)dk_2 - \int e^{-ik_2(x_2'+x_2)} \left(\int e^{2ik_2'x_2} R_{\alpha\beta}(\vec{k}-\vec{k}')R_{\mu\nu}(\vec{k}')d^3\vec{k}'\right)dk_2$

$+ \int e^{-ik_2(x_2'+x_2)} \left(\int e^{2ik_2'(x_2'+x_2)} R_{\alpha\beta}(\vec{k}-\vec{k}')R_{\mu\nu}(\vec{k}')d^3\vec{k}'\right)dk_2 + \int e^{-ik_2(x_2'+x_2)} \left(\int R_{\alpha\beta}(\vec{k}-\vec{k}')R_{\mu\nu}(\vec{k}')d^3\vec{k}'\right)dk_2 ]$

(A-15)

APPENDIX B. Results of the square bracket [ ] in Eq. 12 for 16 cases

Case 1: when $S(x_2'',x_2',\vec{\kappa},\omega)$ is replaced by $\int e^{ik_2(x_2''-x_2')} \left(\int R_{\alpha\beta}(\vec{k}-\vec{k}')R_{\mu\nu}(\vec{k}')d^3\vec{k}'\right)dk_2$,

$$[\ ] = \iint \dfrac{4\kappa^2 + 8\kappa k_2 e^{-\kappa x_2}\sin(k_2 x_2) + 4k_2^2 e^{-2\kappa x_2}}{(\kappa^2 + k_2^2)^2} R_{\alpha\beta}(\vec{k}-\vec{k}')R_{\mu\nu}(\vec{k}')d^3\vec{k}'dk_2 \quad (B-1)$$

Case 2: when $S(x_2'',x_2',\vec{\kappa},\omega)$ is replaced by $\int e^{ik_2(x_2''-x_2')} \left(\int e^{-2ik_2'(x_2''-x_2')} R_{\alpha\beta}(\vec{k}-\vec{k}')R_{\mu\nu}(\vec{k}')d^3\vec{k}'\right)dk_2$,

$$[\ ] = \iint \dfrac{4\kappa^2 + 8\kappa(k_2-2k_2')e^{-\kappa x_2}\sin[(k_2-2k_2')x_2] + 4(k_2-2k_2')^2 e^{-2\kappa x_2}}{[\kappa^2 + (k_2-2k_2')^2]^2} R_{\alpha\beta}(\vec{k}-\vec{k}')R_{\mu\nu}(\vec{k}')d^3\vec{k}'dk_2$$

(B-2)

Case 3: when $S(x_2'',x_2',\vec{\kappa},\omega)$ is replaced by $\int e^{ik_2(x_2''-x_2')} \left(\int e^{2ik_2'x_2'} R_{\alpha\beta}(\vec{k}-\vec{k}')R_{\mu\nu}(\vec{k}')d^3\vec{k}'\right)dk_2$,

$$[\ ] = \iint \dfrac{4\kappa^2 e^{2ik_2'x_2} + 4\kappa k_2 i e^{(-ik_2+2ik_2'-\kappa)x_2} - 4\kappa(k_2-2k_2')ie^{(ik_2-\kappa)x_2} + 4k_2(k_2-2k_2')e^{-2\kappa x_2}}{[\kappa^2+(k_2-2k_2')^2](\kappa^2+k_2^2)} R_{\alpha\beta}(\vec{k}-\vec{k}')R_{\mu\nu}(\vec{k}')d^3\vec{k}'dk_2$$

(B-3)

Case 4: when $S(x_2'',x_2',\vec{\kappa},\omega)$ is replaced by $\int e^{ik_2(x_2''-x_2')} \left(\int e^{-2ik_2'x_2''} R_{\alpha\beta}(\vec{k}-\vec{k}')R_{\mu\nu}(\vec{k}')d^3\vec{k}'\right)dk_2$,

$$[\ ] = \iint \dfrac{4\kappa^2 e^{-2ik_2'x_2} - 4\kappa k_2 i e^{(ik_2-2ik_2'-\kappa)x_2} + 4\kappa(k_2-2k_2')ie^{(-ik_2-\kappa)x_2} + 4k_2(k_2-2k_2')e^{-2\kappa x_2}}{[\kappa^2+(k_2-2k_2')^2](\kappa^2+k_2^2)} R_{\alpha\beta}(\vec{k}-\vec{k}')R_{\mu\nu}(\vec{k}')d^3\vec{k}'dk_2$$

(B-4)

Case 5: when $S(x_2'',x_2',\vec{\kappa},\omega)$ is replaced by $\int e^{-ik_2(x_2''-x_2')} \left(\int e^{2ik_2'(x_2''-x_2')} R_{\alpha\beta}(\vec{k}-\vec{k}')R_{\mu\nu}(\vec{k}')d^3\vec{k}'\right)dk_2$,

$$[\ ] = \iint \dfrac{4\kappa^2 + 8\kappa(k_2-2k_2')e^{-\kappa x_2}\sin[(k_2-2k_2')x_2] + 4(k_2-2k_2')^2 e^{-2\kappa x_2}}{[\kappa^2+(k_2-2k_2')^2]^2} R_{\alpha\beta}(\vec{k}-\vec{k}')R_{\mu\nu}(\vec{k}')d^3\vec{k}'dk_2 \quad (B-5)$$

Case 6: when $S(x_2",x_2',\vec{\kappa},\omega)$ is replaced by $\int e^{-ik_2(x_2"-x_2')}\left(\int R_{\alpha\beta}(\vec{k}-\vec{k}')R_{\mu\nu}(\vec{k}')d^3\vec{k}'\right)dk_2$,

$$[\ ]=\iint\frac{4\kappa^2+8\kappa k_2 e^{-\kappa x_2}\sin(k_2 x_2)+4k_2^2 e^{-2\kappa x_2}}{(\kappa^2+k_2^2)^2}R_{\alpha\beta}(\vec{k}-\vec{k}')R_{\mu\nu}(\vec{k}')d^3\vec{k}'dk_2 \qquad \text{(B-6)}$$

Case 7: when $S(x_2",x_2',\vec{\kappa},\omega)$ is replaced by $\int e^{-ik_2(x_2"-x_2')}\left(\int e^{2ik_2'x_2"}R_{\alpha\beta}(\vec{k}-\vec{k}')R_{\mu\nu}(\vec{k}')d^3\vec{k}'\right)dk_2$,

$$[\ ]=\iint\frac{4\kappa^2 e^{2ik_2'x_2}+4\kappa k_2 ie^{(-ik_2+2ik_2'-\kappa)x_2}-4\kappa(k_2-2k_2')ie^{(ik_2-\kappa)x_2}+4k_2(k_2-2k_2')e^{-2\kappa x_2}}{[\kappa^2+(k_2-2k_2')^2](\kappa^2+k_2^2)}R_{\alpha\beta}(\vec{k}-\vec{k}')R_{\mu\nu}(\vec{k}')d^3\vec{k}'dk_2$$

(B-7)

Case 8: when $S(x_2",x_2',\vec{\kappa},\omega)$ is replaced by $\int e^{-ik_2(x_2"-x_2')}\left(\int e^{-2ik_2'x_2'}R_{\alpha\beta}(\vec{k}-\vec{k}')R_{\mu\nu}(\vec{k}')d^3\vec{k}'\right)dk_2$,

$$[\ ]=\iint\frac{4\kappa^2 e^{-2ik_2'x_2}-4\kappa k_2 ie^{(ik_2-2ik_2'-\kappa)x_2}+4\kappa(k_2-2k_2')ie^{(-ik_2-\kappa)x_2}+4k_2(k_2-2k_2')e^{-2\kappa x_2}}{[\kappa^2+(k_2-2k_2')^2](\kappa^2+k_2^2)}R_{\alpha\beta}(\vec{k}-\vec{k}')R_{\mu\nu}(\vec{k}')d^3\vec{k}'dk_2$$

(B-8)

Case 9: when $S(x_2",x_2',\vec{\kappa},\omega)$ is replaced by $\int e^{ik_2(x_2"+x_2')}\left(\int e^{-2ik_2'x_2'}R_{\alpha\beta}(\vec{k}-\vec{k}')R_{\mu\nu}(\vec{k}')d^3\vec{k}'\right)dk_2$,

$$[\ ]=\iint\frac{4\kappa^2 e^{(2ik_2-2ik_2')x_2}+4\kappa k_2 ie^{(ik_2-2ik_2'-\kappa)x_2}+4\kappa(k_2-2k_2')ie^{(ik_2-\kappa)x_2}-4k_2(k_2-2k_2')e^{-2\kappa x_2}}{[\kappa^2+(k_2-2k_2')^2](\kappa^2+k_2^2)}R_{\alpha\beta}(\vec{k}-\vec{k}')R_{\mu\nu}(\vec{k}')d^3\vec{k}'dk_2$$

(B-9)

Case 10: when $S(x_2",x_2',\vec{\kappa},\omega)$ is replaced by $\int e^{ik_2(x_2"+x_2')}\left(\int e^{-2ik_2'x_2"}R_{\alpha\beta}(\vec{k}-\vec{k}')R_{\mu\nu}(\vec{k}')d^3\vec{k}'\right)dk_2$,

$$[\ ]=\iint\frac{4\kappa^2 e^{(2ik_2-2ik_2')x_2}+4\kappa k_2 ie^{(ik_2-2ik_2'-\kappa)x_2}+4\kappa(k_2-2k_2')ie^{(ik_2-\kappa)x_2}-4k_2(k_2-2k_2')e^{-2\kappa x_2}}{[\kappa^2+(k_2-2k_2')^2](\kappa^2+k_2^2)}R_{\alpha\beta}(\vec{k}-\vec{k}')R_{\mu\nu}(\vec{k}')d^3\vec{k}'dk_2$$

(B-10)

Case 11: when $S(x_2",x_2',\vec{\kappa},\omega)$ is replaced by $\int e^{ik_2(x_2"+x_2')}\left(\int R_{\alpha\beta}(\vec{k}-\vec{k}')R_{\mu\nu}(\vec{k}')d^3\vec{k}'\right)dk_2$,

$$[\ ]=\iint\frac{4\kappa^2 e^{2ik_2 x_2}+8\kappa k_2 ie^{(ik_2-\kappa)x_2}-4k_2^2 e^{-2\kappa x_2}}{(\kappa^2+k_2^2)^2}R_{\alpha\beta}(\vec{k}-\vec{k}')R_{\mu\nu}(\vec{k}')d^3\vec{k}'dk_2 \qquad \text{(B-11)}$$

Case 12: when $S(x_2",x_2',\vec{\kappa},\omega)$ is replaced by $\int e^{ik_2(x_2"+x_2')}\left(\int e^{-2ik_2'(x_2"+x_2')}R_{\alpha\beta}(\vec{k}-\vec{k}')R_{\mu\nu}(\vec{k}')d^3\vec{k}'\right)dk_2$,

$$[\ ]=\iint\frac{4\kappa^2 e^{(2ik_2-4ik_2')x_2}+8\kappa(k_2-2k_2')ie^{(ik_2-2ik_2'-\kappa)x_2}-4(k_2-2k_2')^2 e^{-2\kappa x_2}}{[\kappa^2+(k_2-2k_2')^2]^2}R_{\alpha\beta}(\vec{k}-\vec{k}')R_{\mu\nu}(\vec{k}')d^3\vec{k}'dk_2$$

(B-12)

Case 13: when $S(x_2",x_2',\vec{\kappa},\omega)$ is replaced by $\int e^{-ik_2(x_2"+x_2')}\left(\int e^{2ik_2'x_2"}R_{\alpha\beta}(\vec{k}-\vec{k}')R_{\mu\nu}(\vec{k}')d^3\vec{k}'\right)dk_2$,

$$[\ ]=\iint\frac{4\kappa^2 e^{(-2ik_2+2ik_2')x_2}-4\kappa k_2 ie^{(-ik_2+2ik_2'-\kappa)x_2}-4\kappa(k_2-2k_2')ie^{(-ik_2-\kappa)x_2}-4k_2(k_2-2k_2')e^{-2\kappa x_2}}{[\kappa^2+(k_2-2k_2')^2](\kappa^2+k_2^2)}R_{\alpha\beta}(\vec{k}-\vec{k}')R_{\mu\nu}(\vec{k}')d^3\vec{k}'dk_2$$

(B-13)

Case 14: when $S(x_2",x_2',\vec{\kappa},\omega)$ is replaced by $\int e^{-ik_2(x_2"+x_2')}\left(\int e^{2ik_2'x_2'}R_{\alpha\beta}(\vec{k}-\vec{k}')R_{\mu\nu}(\vec{k}')d^3\vec{k}'\right)dk_2$,

$$[\ ] = \iint \frac{4\kappa^2 e^{(-2ik_2+2ik_2')x_2} - 4\kappa k_2' i e^{(-ik_2+2ik_2'-\kappa)x_2} - 4\kappa(k_2-2k_2')ie^{(-ik_2-\kappa)x_2} - 4k_2(k_2-2k_2')e^{-2\kappa x_2}}{[\kappa^2+(k_2-2k_2')^2](\kappa^2+k_2^2)} R_{\alpha\beta}(\vec{k}-\vec{k}')R_{\mu\nu}(\vec{k}')d^3\vec{k}'dk_2$$

(B-14)

Case 15: when $S(x_2",x_2',\vec{\kappa},\omega)$ is replaced by $\int e^{-ik_2(x_2"+x_2')}\left(\int e^{2ik_2'(x_2"+x_2')}R_{\alpha\beta}(\vec{k}-\vec{k}')R_{\mu\nu}(\vec{k}')d^3\vec{k}'\right)dk_2$,

$$[\ ] = \iint \frac{4\kappa^2 e^{(-2ik_2+4ik_2')x_2} - 8\kappa(k_2-2k_2')ie^{(-ik_2+2ik_2'-\kappa)x_2} - 4(k_2-2k_2')^2 e^{-2\kappa x_2}}{[\kappa^2+(k_2-2k_2')^2]^2} R_{\alpha\beta}(\vec{k}-\vec{k}')R_{\mu\nu}(\vec{k}')d^3\vec{k}'dk_2$$

(B-15)

Case 16: when $S(x_2",x_2',\vec{\kappa},\omega)$ is replaced by $\int e^{-ik_2(x_2"+x_2')}\left(\int R_{\alpha\beta}(\vec{k}-\vec{k}')R_{\mu\nu}(\vec{k}')d^3\vec{k}'\right)dk_2$,

$$[\ ] = \iint \frac{4\kappa^2 e^{-2ik_2 x_2} - 8\kappa k_2 i e^{(-ik_2-\kappa)x_2} - 4k_2^2 e^{-2\kappa x_2}}{(\kappa^2+k_2^2)^2} R_{\alpha\beta}(\vec{k}-\vec{k}')R_{\mu\nu}(\vec{k}')d^3\vec{k}'dk_2 \qquad \text{(B-16)}$$

## APPENDIX C. Form of $|p(x_2,k_1)|^2$

Note that $\int dA$ refers to $\int_{-\infty}^{\infty}\int_{-\infty}^{\infty}\int_{-\infty}^{\infty}\int_{-\infty}^{\infty}\int_{-\infty}^{\infty} dk_1'dk_2'dk_3'dk_2dk_3$ for simplicity.

$$\begin{aligned}
&|p(x_2,k_1)|^2 \\
&= (2\pi)^{-2}\left(\frac{55C}{18}\right)^2 \frac{\rho^2\lambda^8}{2}\Big\{\Big[\int \frac{2k_1^4}{\kappa^2}\frac{\kappa^2+2\kappa k_2 e^{-\kappa x_2}\sin(k_2 x_2)+k_2^2 e^{-2\kappa x_2}}{(\kappa^2+k_2^2)^2[1+|\vec{k}-\vec{k}'|^2\lambda^2]^{17/6}(1+k'^2\lambda^2)^{17/6}}[(k_2-k_2')^2+(k_3-k_3')^2](k_2'^2+k_3'^2)dA \\
&+\int \frac{2k_3^4}{\kappa^2}\frac{\kappa^2+2\kappa k_2 e^{-\kappa x_2}\sin(k_2 x_2)+k_2^2 e^{-2\kappa x_2}}{(\kappa^2+k_2^2)^2[1+|\vec{k}-\vec{k}'|^2\lambda^2]^{17/6}(1+k'^2\lambda^2)^{17/6}}[(k_1-k_1')^2+(k_2-k_2')^2](k_1'^2+k_2'^2)dA \\
&+\int \frac{2k_1^2 k_3^2}{\kappa^2}\frac{\kappa^2+2\kappa k_2 e^{-\kappa x_2}\sin(k_2 x_2)+k_2^2 e^{-2\kappa x_2}}{(\kappa^2+k_2^2)^2[1+|\vec{k}-\vec{k}'|^2\lambda^2]^{17/6}(1+k'^2\lambda^2)^{17/6}}[(k_2-k_2')^2+(k_3-k_3')^2](k_1'^2+k_2'^2)dA \\
&+\int \frac{2k_1^2 k_3^2}{\kappa^2}\frac{\kappa^2+2\kappa k_2 e^{-\kappa x_2}\sin(k_2 x_2)+k_2^2 e^{-2\kappa x_2}}{(\kappa^2+k_2^2)^2[1+|\vec{k}-\vec{k}'|^2\lambda^2]^{17/6}(1+k'^2\lambda^2)^{17/6}}[(k_1-k_1')^2+(k_2-k_2')^2](k_2'^2+k_3'^2)dA \\
&-\int \frac{4k_1^3 k_3}{\kappa^2}\frac{\kappa^2+2\kappa k_2 e^{-\kappa x_2}\sin(k_2 x_2)+k_2^2 e^{-2\kappa x_2}}{(\kappa^2+k_2^2)^2[1+|\vec{k}-\vec{k}'|^2\lambda^2]^{17/6}(1+k'^2\lambda^2)^{17/6}}[(k_2-k_2')^2+(k_3-k_3')^2]k_1'k_3'dA \\
&-\int \frac{4k_1^3 k_3}{\kappa^2}\frac{\kappa^2+2\kappa k_2 e^{-\kappa x_2}\sin(k_2 x_2)+k_2^2 e^{-2\kappa x_2}}{(\kappa^2+k_2^2)^2[1+|\vec{k}-\vec{k}'|^2\lambda^2]^{17/6}(1+k'^2\lambda^2)^{17/6}}(k_1-k_1')(k_3-k_3')(k_2'^2+k_3'^2)dA \\
&-\int \frac{4k_1 k_3^3}{\kappa^2}\frac{\kappa^2+2\kappa k_2 e^{-\kappa x_2}\sin(k_2 x_2)+k_2^2 e^{-2\kappa x_2}}{(\kappa^2+k_2^2)^2[1+|\vec{k}-\vec{k}'|^2\lambda^2]^{17/6}(1+k'^2\lambda^2)^{17/6}}[(k_1-k_1')^2+(k_2-k_2')^2]k_1'k_3'dA \\
&-\int \frac{4k_1 k_3^3}{\kappa^2}\frac{\kappa^2+2\kappa k_2 e^{-\kappa x_2}\sin(k_2 x_2)+k_2^2 e^{-2\kappa x_2}}{(\kappa^2+k_2^2)^2[1+|\vec{k}-\vec{k}'|^2\lambda^2]^{17/6}(1+k'^2\lambda^2)^{17/6}}(k_1-k_1')(k_3-k_3')(k_1'^2+k_2'^2)dA \\
&+\int \frac{8k_1^2 k_3^2}{\kappa^2}\frac{\kappa^2+2\kappa k_2 e^{-\kappa x_2}\sin(k_2 x_2)+k_2^2 e^{-2\kappa x_2}}{(\kappa^2+k_2^2)^2[1+|\vec{k}-\vec{k}'|^2\lambda^2]^{17/6}(1+k'^2\lambda^2)^{17/6}}(k_1-k_1')(k_3-k_3')k_1'k_3'dA\Big] \\
&-\Big[\int \frac{4k_1^2 k_3 k_2}{\sqrt{2}\kappa^2}\frac{\kappa^2+2\kappa k_2 e^{-\kappa x_2}\sin(k_2 x_2)+k_2^2 e^{-2\kappa x_2}}{(\kappa^2+k_2^2)^2[1+|\vec{k}-\vec{k}'|^2\lambda^2]^{17/6}(1+k'^2\lambda^2)^{17/6}}[(k_2-k_2')^2+(k_3-k_3')^2](k_2'k_3')dA \\
&+\int \frac{4k_1^2 k_3(k_2-2k_2')}{\sqrt{2}\kappa^2}\frac{\kappa^2\cos(2k_2'x_2)+\kappa k_2 e^{-\kappa x_2}\sin[(k_2-2k_2')x_2]+\kappa(k_2-2k_2')e^{-\kappa x_2}\sin(k_2 x_2)+k_2(k_2-2k_2')e^{-2\kappa x_2}}{[\kappa^2+(k_2-2k_2')^2](\kappa^2+k_2^2)[1+|\vec{k}-\vec{k}'|^2\lambda^2]^{17/6}(1+k'^2\lambda^2)^{17/6}}[(k_2-k_2')^2+(k_3-k_3')^2](k_2'k_3')dA
\end{aligned}$$

(C-1)

$$-\int \frac{4k_1^2 k_3 k_2}{\sqrt{2}\kappa^2} \frac{\kappa^2 + 2\kappa k_2 e^{-\kappa x_2}\sin(k_2 x_2) + k_2^2 e^{-2\kappa x_2}}{(\kappa^2 + k_2^2)^2 [1+|\vec{k}-\vec{k}'|^2 \lambda^2]^{17/6}(1+k'^2\lambda^2)^{17/6}} (k_2-k_2')(k_3-k_3')(k_2'^2+k_3'^2) dA$$

$$-\int \frac{4k_1^2 k_3 (k_2-2k_2')}{\sqrt{2}\kappa^2} \frac{\kappa^2 \cos[2(k_2-k_2')x_2] - \kappa k_2 e^{-\kappa x_2}\sin[(k_2-2k_2')x_2] - \kappa(k_2-2k_2')e^{-\kappa x_2}\sin(k_2 x_2) - k_2(k_2-2k_2')e^{-2\kappa x_2}}{[\kappa^2+(k_2-2k_2')^2](\kappa^2+k_2^2)[1+|\vec{k}-\vec{k}'|^2\lambda^2]^{17/6}(1+k'^2\lambda^2)^{17/6}} (k_2-k_2')(k_3-k_3')(k_2'^2+k_3'^2) dA$$

$$-\int \frac{4k_3^3 k_2}{\sqrt{2}\kappa^2} \frac{\kappa^2 + 2\kappa k_2 e^{-\kappa x_2}\sin(k_2 x_2) + k_2^2 e^{-2\kappa x_2}}{(\kappa^2+k_2^2)^2 [1+|\vec{k}-\vec{k}'|^2\lambda^2]^{17/6}(1+k'^2\lambda^2)^{17/6}} [(k_1-k_1')^2+(k_2-k_2')^2] k_2' k_3' dA$$

$$+\int \frac{4k_3^3 (k_2-2k_2')}{\sqrt{2}\kappa^2} \frac{\kappa^2 \cos(2k_2' x_2) + \kappa k_2 e^{-\kappa x_2}\sin[(k_2-2k_2')x_2] + \kappa(k_2-2k_2')e^{-\kappa x_2}\sin(k_2 x_2) + k_2(k_2-2k_2')e^{-2\kappa x_2}}{[\kappa^2+(k_2-2k_2')^2](\kappa^2+k_2^2)[1+|\vec{k}-\vec{k}'|^2\lambda^2]^{17/6}(1+k'^2\lambda^2)^{17/6}} [(k_1-k_1')^2+(k_2-k_2')^2] k_2' k_3' dA$$

$$-\int \frac{4k_3^3 k_2}{\sqrt{2}\kappa^2} \frac{\kappa^2 + 2\kappa k_2 e^{-\kappa x_2}\sin(k_2 x_2) + k_2^2 e^{-2\kappa x_2}}{(\kappa^2+k_2^2)^2 [1+|\vec{k}-\vec{k}'|^2\lambda^2]^{17/6}(1+k'^2\lambda^2)^{17/6}} (k_2-k_2')(k_3-k_3')(k_1'^2+k_2'^2) dA$$

$$-\int \frac{4k_3^3 (k_2-2k_2')}{\sqrt{2}\kappa^2} \frac{\kappa^2 \cos[2(k_2-k_2')x_2] - \kappa k_2 e^{-\kappa x_2}\sin[(k_2-2k_2')x_2] - \kappa(k_2-2k_2')e^{-\kappa x_2}\sin(k_2 x_2) - k_2(k_2-2k_2')e^{-2\kappa x_2}}{[\kappa^2+(k_2-2k_2')^2](\kappa^2+k_2^2)[1+|\vec{k}-\vec{k}'|^2\lambda^2]^{17/6}(1+k'^2\lambda^2)^{17/6}} (k_2-k_2')(k_3-k_3')(k_1'^2+k_2'^2) dA$$

$$-\int \frac{4k_1^3 k_2}{\sqrt{2}\kappa^2} \frac{\kappa^2 + 2\kappa k_2 e^{-\kappa x_2}\sin(k_2 x_2) + k_2^2 e^{-2\kappa x_2}}{(\kappa^2+k_2^2)^2 [1+|\vec{k}-\vec{k}'|^2\lambda^2]^{17/6}(1+k'^2\lambda^2)^{17/6}} [(k_2-k_2')^2+(k_3-k_3')^2] k_2' k_1' dA$$

$$+\int \frac{4k_1^3 (k_2-2k_2')}{\sqrt{2}\kappa^2} \frac{\kappa^2 \cos(2k_2' x_2) + \kappa k_2 e^{-\kappa x_2}\sin[(k_2-2k_2')x_2] + \kappa(k_2-2k_2')e^{-\kappa x_2}\sin(k_2 x_2) + k_2(k_2-2k_2')e^{-2\kappa x_2}}{[\kappa^2+(k_2-2k_2')^2](\kappa^2+k_2^2)[1+|\vec{k}-\vec{k}'|^2\lambda^2]^{17/6}(1+k'^2\lambda^2)^{17/6}} [(k_2-k_2')^2+(k_3-k_3')^2] k_2' k_1' dA$$

$$-\int \frac{4k_1^3 k_2}{\sqrt{2}\kappa^2} \frac{\kappa^2 + 2\kappa k_2 e^{-\kappa x_2}\sin(k_2 x_2) + k_2^2 e^{-2\kappa x_2}}{(\kappa^2+k_2^2)^2 [1+|\vec{k}-\vec{k}'|^2\lambda^2]^{17/6}(1+k'^2\lambda^2)^{17/6}} (k_2-k_2')(k_1-k_1')(k_2'^2+k_3'^2) dA$$

$$-\int \frac{4k_1^3 (k_2-2k_2')}{\sqrt{2}\kappa^2} \frac{\kappa^2 \cos[2(k_2-k_2')x_2] - \kappa k_2 e^{-\kappa x_2}\sin[(k_2-2k_2')x_2] - \kappa(k_2-2k_2')e^{-\kappa x_2}\sin(k_2 x_2) - k_2(k_2-2k_2')e^{-2\kappa x_2}}{[\kappa^2+(k_2-2k_2')^2](\kappa^2+k_2^2)[1+|\vec{k}-\vec{k}'|^2\lambda^2]^{17/6}(1+k'^2\lambda^2)^{17/6}} (k_2-k_2')(k_1-k_1')(k_2'^2+k_3'^2) dA$$

$$-\int \frac{4k_1 k_3^2 k_2}{\sqrt{2}\kappa^2} \frac{\kappa^2 + 2\kappa k_2 e^{-\kappa x_2}\sin(k_2 x_2) + k_2^2 e^{-2\kappa x_2}}{(\kappa^2+k_2^2)^2 [1+|\vec{k}-\vec{k}'|^2\lambda^2]^{17/6}(1+k'^2\lambda^2)^{17/6}} [(k_1-k_1')^2+(k_2-k_2')^2] k_2' k_1' dA$$

$$+\int \frac{4k_1 k_3^2 (k_2-2k_2')}{\sqrt{2}\kappa^2} \frac{\kappa^2 \cos(2k_2' x_2) + \kappa k_2 e^{-\kappa x_2}\sin[(k_2-2k_2')x_2] + \kappa(k_2-2k_2')e^{-\kappa x_2}\sin(k_2 x_2) + k_2(k_2-2k_2')e^{-2\kappa x_2}}{[\kappa^2+(k_2-2k_2')^2](\kappa^2+k_2^2)[1+|\vec{k}-\vec{k}'|^2\lambda^2]^{17/6}(1+k'^2\lambda^2)^{17/6}} [(k_1-k_1')^2+(k_2-k_2')^2] k_2' k_1' dA$$

$$-\int \frac{4k_1 k_3^2 k_2}{\sqrt{2}\kappa^2} \frac{\kappa^2 + 2\kappa k_2 e^{-\kappa x_2}\sin(k_2 x_2) + k_2^2 e^{-2\kappa x_2}}{(\kappa^2+k_2^2)^2 [1+|\vec{k}-\vec{k}'|^2\lambda^2]^{17/6}(1+k'^2\lambda^2)^{17/6}} (k_2-k_2')(k_1-k_1')(k_1'^2+k_2'^2) dA$$

$$-\int \frac{4k_1 k_3^2 (k_2-2k_2')}{\sqrt{2}\kappa^2} \frac{\kappa^2 \cos[2(k_2-k_2')x_2] - \kappa k_2 e^{-\kappa x_2}\sin[(k_2-2k_2')x_2] - \kappa(k_2-2k_2')e^{-\kappa x_2}\sin(k_2 x_2) - k_2(k_2-2k_2')e^{-2\kappa x_2}}{[\kappa^2+(k_2-2k_2')^2](\kappa^2+k_2^2)[1+|\vec{k}-\vec{k}'|^2\lambda^2]^{17/6}(1+k'^2\lambda^2)^{17/6}} (k_2-k_2')(k_1-k_1')(k_1'^2+k_2'^2) dA$$

$$+\int \frac{8k_1^2 k_3 k_2}{\sqrt{2}\kappa^2} \frac{\kappa^2 + 2\kappa k_2 e^{-\kappa x_2}\sin(k_2 x_2) + k_2^2 e^{-2\kappa x_2}}{(\kappa^2+k_2^2)^2 [1+|\vec{k}-\vec{k}'|^2\lambda^2]^{17/6}(1+k'^2\lambda^2)^{17/6}} (k_1-k_1')(k_3-k_3') k_2' k_1' dA$$

$$-\int \frac{8k_1^2 k_3 (k_2-2k_2')}{\sqrt{2}\kappa^2} \frac{\kappa^2 \cos(2k_2' x_2) + \kappa k_2 e^{-\kappa x_2}\sin[(k_2-2k_2')x_2] + \kappa(k_2-2k_2')e^{-\kappa x_2}\sin(k_2 x_2) + k_2(k_2-2k_2')e^{-2\kappa x_2}}{[\kappa^2+(k_2-2k_2')^2](\kappa^2+k_2^2)[1+|\vec{k}-\vec{k}'|^2\lambda^2]^{17/6}(1+k'^2\lambda^2)^{17/6}} (k_1-k_1')(k_3-k_3') k_2' k_1' dA$$

$$+\int \frac{8k_1^2 k_3 k_2}{\sqrt{2}\kappa^2} \frac{\kappa^2 + 2\kappa k_2 e^{-\kappa x_2}\sin(k_2 x_2) + k_2^2 e^{-2\kappa x_2}}{(\kappa^2+k_2^2)^2 [1+|\vec{k}-\vec{k}'|^2\lambda^2]^{17/6}(1+k'^2\lambda^2)^{17/6}} (k_2-k_2')(k_1-k_1') k_1' k_3' dA$$

$$+\int \frac{8k_1^2 k_3 (k_2-2k_2')}{\sqrt{2}\kappa^2} \frac{\kappa^2 \cos[2(k_2-k_2')x_2] - \kappa k_2 e^{-\kappa x_2}\sin[(k_2-2k_2')x_2] - \kappa(k_2-2k_2')e^{-\kappa x_2}\sin(k_2 x_2) - k_2(k_2-2k_2')e^{-2\kappa x_2}}{[\kappa^2+(k_2-2k_2')^2](\kappa^2+k_2^2)[1+|\vec{k}-\vec{k}'|^2\lambda^2]^{17/6}(1+k'^2\lambda^2)^{17/6}} (k_2-k_2')(k_1-k_1') k_1' k_3' dA$$

$$+\int \frac{8k_1 k_3^2 k_2}{\sqrt{2}\kappa^2} \frac{\kappa^2 + 2\kappa k_2 e^{-\kappa x_2}\sin(k_2 x_2) + k_2^2 e^{-2\kappa x_2}}{(\kappa^2+k_2^2)^2 [1+|\vec{k}-\vec{k}'|^2\lambda^2]^{17/6}(1+k'^2\lambda^2)^{17/6}} (k_1-k_1')(k_3-k_3') k_2' k_3' dA$$

$$-\int \frac{8k_1 k_3^2 (k_2-2k_2')}{\sqrt{2}\kappa^2} \frac{\kappa^2 \cos(2k_2' x_2) + \kappa k_2 e^{-\kappa x_2}\sin[(k_2-2k_2')x_2] + \kappa(k_2-2k_2')e^{-\kappa x_2}\sin(k_2 x_2) + k_2(k_2-2k_2')e^{-2\kappa x_2}}{[\kappa^2+(k_2-2k_2')^2](\kappa^2+k_2^2)[1+|\vec{k}-\vec{k}'|^2\lambda^2]^{17/6}(1+k'^2\lambda^2)^{17/6}} (k_1-k_1')(k_3-k_3') k_2' k_3' dA$$

$$+\int \frac{8k_1 k_3^2 k_2}{\sqrt{2}\kappa^2} \frac{\kappa^2 + 2\kappa k_2 e^{-\kappa x_2}\sin(k_2 x_2) + k_2^2 e^{-2\kappa x_2}}{(\kappa^2+k_2^2)^2 [1+|\vec{k}-\vec{k}'|^2\lambda^2]^{17/6}(1+k'^2\lambda^2)^{17/6}} (k_2-k_2')(k_3-k_3') k_1' k_3' dA$$

$$+\int \frac{8k_1 k_3^2 (k_2-2k_2')}{\sqrt{2}\kappa^2} \frac{\kappa^2 \cos[2(k_2-k_2')x_2] - \kappa k_2 e^{-\kappa x_2}\sin[(k_2-2k_2')x_2] - \kappa(k_2-2k_2')e^{-\kappa x_2}\sin(k_2 x_2) - k_2(k_2-2k_2')e^{-2\kappa x_2}}{[\kappa^2+(k_2-2k_2')^2](\kappa^2+k_2^2)[1+|\vec{k}-\vec{k}'|^2\lambda^2]^{17/6}(1+k'^2\lambda^2)^{17/6}} (k_2-k_2')(k_3-k_3') k_1' k_3' dA]$$

$$+[\int \frac{k_1^2 k_2^2}{\kappa^2} \frac{\kappa^2 + 2\kappa k_2 e^{-\kappa x_2}\sin(k_2 x_2) + k_2^2 e^{-2\kappa x_2}}{(\kappa^2+k_2^2)^2 [1+|\vec{k}-\vec{k}'|^2\lambda^2]^{17/6}(1+k'^2\lambda^2)^{17/6}} [(k_2-k_2')^2+(k_3-k_3')^2](k_1'^2+k_3'^2) dA$$

$$+\int \frac{k_1^2 (k_2-2k_2')^2}{\kappa^2} \frac{\kappa^2 + 2\kappa (k_2-2k_2')e^{-\kappa x_2}\sin[(k_2-2k_2')x_2] + (k_2-2k_2')^2 e^{-2\kappa x_2}}{[\kappa^2+(k_2-2k_2')^2]^2 [1+|\vec{k}-\vec{k}'|^2\lambda^2]^{17/6}(1+k'^2\lambda^2)^{17/6}} [(k_2-k_2')^2+(k_3-k_3')^2](k_1'^2+k_3'^2) dA$$

$$-\int \frac{2k_1^2 k_2 (k_2-2k_2')}{\kappa^2} \frac{\kappa^2 \cos(2k_2' x_2) + \kappa k_2 e^{-\kappa x_2}\sin[(k_2-2k_2')x_2] + \kappa(k_2-2k_2')e^{-\kappa x_2}\sin(k_2 x_2) + k_2(k_2-2k_2')e^{-2\kappa x_2}}{[\kappa^2+(k_2-2k_2')^2](\kappa^2+k_2^2)[1+|\vec{k}-\vec{k}'|^2\lambda^2]^{17/6}(1+k'^2\lambda^2)^{17/6}} [(k_2-k_2')^2+(k_3-k_3')^2](k_1'^2+k_3'^2) dA$$

$$+ \int \frac{k_1^2 k_2^2}{\kappa^2} \frac{\kappa^2 + 2\kappa k_2 e^{-\kappa x_2}\sin(k_2 x_2) + k_2^2 e^{-2\kappa x_2}}{(\kappa^2 + k_2^2)^2 [1 + |\vec{k} - \vec{k}'|^2 \lambda^2]^{17/6} (1 + k'^2 \lambda^2)^{17/6}} [(k_1 - k_1')^2 + (k_3 - k_3')^2](k_2'^2 + k_3'^2) dA$$

$$+ \int \frac{k_1^2 (k_2 - 2k_2')^2}{\kappa^2} \frac{\kappa^2 + 2\kappa (k_2 - 2k_2')e^{-\kappa x_2}\sin[(k_2 - 2k_2')x_2] + (k_2 - 2k_2')^2 e^{-2\kappa x_2}}{[\kappa^2 + (k_2 - 2k_2')^2]^2 [1 + |\vec{k} - \vec{k}'|^2 \lambda^2]^{17/6} (1 + k'^2 \lambda^2)^{17/6}} [(k_1 - k_1')^2 + (k_3 - k_3')^2](k_2'^2 + k_3'^2) dA$$

$$+ \int \frac{2k_1^2 k_2 (k_2 - 2k_2')}{\kappa^2} \frac{\kappa^2 \cos[2(k_2 - k_2')x_2] - \kappa k_2 e^{-\kappa x_2}\sin[(k_2 - 2k_2')x_2] - \kappa(k_2 - 2k_2')e^{-\kappa x_2}\sin(k_2 x_2) - k_2(k_2 - 2k_2')e^{-2\kappa x_2}}{[\kappa^2 + (k_2 - 2k_2')^2](\kappa^2 + k_2^2)[1 + |\vec{k} - \vec{k}'|^2 \lambda^2]^{17/6} (1 + k'^2 \lambda^2)^{17/6}} [(k_1 - k_1')^2 + (k_3 - k_3')^2](k_2'^2 + k_3'^2) dA$$

$$+ \int \frac{k_3^2 k_2^2}{\kappa^2} \frac{\kappa^2 + 2\kappa k_2 e^{-\kappa x_2}\sin(k_2 x_2) + k_2^2 e^{-2\kappa x_2}}{(\kappa^2 + k_2^2)^2 [1 + |\vec{k} - \vec{k}'|^2 \lambda^2]^{17/6} (1 + k'^2 \lambda^2)^{17/6}} [(k_1 - k_1')^2 + (k_2 - k_2')^2](k_1'^2 + k_3'^2) dA$$

$$+ \int \frac{k_3^2 (k_2 - 2k_2')^2}{\kappa^2} \frac{\kappa^2 + 2\kappa (k_2 - 2k_2')e^{-\kappa x_2}\sin[(k_2 - 2k_2')x_2] + (k_2 - 2k_2')^2 e^{-2\kappa x_2}}{[\kappa^2 + (k_2 - 2k_2')^2]^2 [1 + |\vec{k} - \vec{k}'|^2 \lambda^2]^{17/6} (1 + k'^2 \lambda^2)^{17/6}} [(k_1 - k_1')^2 + (k_2 - k_2')^2](k_1'^2 + k_3'^2) dA$$

$$- \int \frac{2k_3^2 k_2 (k_2 - 2k_2')}{\kappa^2} \frac{\kappa^2 \cos(2k_2' x_2) + \kappa k_2 e^{-\kappa x_2}\sin[(k_2 - 2k_2')x_2] + \kappa(k_2 - 2k_2')e^{-\kappa x_2}\sin(k_2 x_2) + k_2(k_2 - 2k_2')e^{-2\kappa x_2}}{[\kappa^2 + (k_2 - 2k_2')^2](\kappa^2 + k_2^2)[1 + |\vec{k} - \vec{k}'|^2 \lambda^2]^{17/6} (1 + k'^2 \lambda^2)^{17/6}} [(k_1 - k_1')^2 + (k_2 - k_2')^2](k_1'^2 + k_3'^2) dA$$

$$+ \int \frac{k_3^2 k_2^2}{\kappa^2} \frac{\kappa^2 + 2\kappa k_2 e^{-\kappa x_2}\sin(k_2 x_2) + k_2^2 e^{-2\kappa x_2}}{(\kappa^2 + k_2^2)^2 [1 + |\vec{k} - \vec{k}'|^2 \lambda^2]^{17/6} (1 + k'^2 \lambda^2)^{17/6}} [(k_1 - k_1')^2 + (k_3 - k_3')^2](k_1'^2 + k_2'^2) dA$$

$$+ \int \frac{k_3^2 (k_2 - 2k_2')^2}{\kappa^2} \frac{\kappa^2 + 2\kappa (k_2 - 2k_2')e^{-\kappa x_2}\sin[(k_2 - 2k_2')x_2] + (k_2 - 2k_2')^2 e^{-2\kappa x_2}}{[\kappa^2 + (k_2 - 2k_2')^2]^2 [1 + |\vec{k} - \vec{k}'|^2 \lambda^2]^{17/6} (1 + k'^2 \lambda^2)^{17/6}} [(k_1 - k_1')^2 + (k_3 - k_3')^2](k_1'^2 + k_2'^2) dA$$

$$+ \int \frac{2k_3^2 k_2 (k_2 - 2k_2')}{\kappa^2} \frac{\kappa^2 \cos[2(k_2 - k_2')x_2] - \kappa k_2 e^{-\kappa x_2}\sin[(k_2 - 2k_2')x_2] - \kappa(k_2 - 2k_2')e^{-\kappa x_2}\sin(k_2 x_2) - k_2(k_2 - 2k_2')e^{-2\kappa x_2}}{[\kappa^2 + (k_2 - 2k_2')^2](\kappa^2 + k_2^2)[1 + |\vec{k} - \vec{k}'|^2 \lambda^2]^{17/6} (1 + k'^2 \lambda^2)^{17/6}} [(k_1 - k_1')^2 + (k_3 - k_3')^2](k_1'^2 + k_2'^2) dA$$

$$- \int \frac{2k_1 k_3 k_2^2}{\kappa^2} \frac{\kappa^2 + 2\kappa k_2 e^{-\kappa x_2}\sin(k_2 x_2) + k_2^2 e^{-2\kappa x_2}}{(\kappa^2 + k_2^2)^2 [1 + |\vec{k} - \vec{k}'|^2 \lambda^2]^{17/6} (1 + k'^2 \lambda^2)^{17/6}} (k_1 - k_1')(k_3 - k_3')(k_1'^2 + k_3'^2) dA$$

$$- \int \frac{2k_1 k_3 (k_2 - 2k_2')^2}{\kappa^2} \frac{\kappa^2 + 2\kappa (k_2 - 2k_2')e^{-\kappa x_2}\sin[(k_2 - 2k_2')x_2] + (k_2 - 2k_2')^2 e^{-2\kappa x_2}}{[\kappa^2 + (k_2 - 2k_2')^2]^2 [1 + |\vec{k} - \vec{k}'|^2 \lambda^2]^{17/6} (1 + k'^2 \lambda^2)^{17/6}} (k_1 - k_1')(k_3 - k_3')(k_1'^2 + k_3'^2) dA$$

$$+ \int \frac{4k_1 k_3 k_2 (k_2 - 2k_2')}{\kappa^2} \frac{\kappa^2 \cos(2k_2' x_2) + \kappa k_2 e^{-\kappa x_2}\sin[(k_2 - 2k_2')x_2] + \kappa(k_2 - 2k_2')e^{-\kappa x_2}\sin(k_2 x_2) + k_2(k_2 - 2k_2')e^{-2\kappa x_2}}{[\kappa^2 + (k_2 - 2k_2')^2](\kappa^2 + k_2^2)[1 + |\vec{k} - \vec{k}'|^2 \lambda^2]^{17/6} (1 + k'^2 \lambda^2)^{17/6}} (k_1 - k_1')(k_3 - k_3')(k_1'^2 + k_3'^2) dA$$

$$- \int \frac{2k_1 k_3 k_2^2}{\kappa^2} \frac{\kappa^2 + 2\kappa k_2 e^{-\kappa x_2}\sin(k_2 x_2) + k_2^2 e^{-2\kappa x_2}}{(\kappa^2 + k_2^2)^2 [1 + |\vec{k} - \vec{k}'|^2 \lambda^2]^{17/6} (1 + k'^2 \lambda^2)^{17/6}} [(k_1 - k_1')^2 + (k_3 - k_3')^2] k_1' k_3' dA$$

$$- \int \frac{2k_1 k_3 (k_2 - 2k_2')^2}{\kappa^2} \frac{\kappa^2 + 2\kappa (k_2 - 2k_2')e^{-\kappa x_2}\sin[(k_2 - 2k_2')x_2] + (k_2 - 2k_2')^2 e^{-2\kappa x_2}}{[\kappa^2 + (k_2 - 2k_2')^2]^2 [1 + |\vec{k} - \vec{k}'|^2 \lambda^2]^{17/6} (1 + k'^2 \lambda^2)^{17/6}} [(k_1 - k_1')^2 + (k_3 - k_3')^2] k_1' k_3' dA$$

$$- \int \frac{4k_1 k_3 k_2 (k_2 - 2k_2')}{\kappa^2} \frac{\kappa^2 \cos[2(k_2 - k_2')x_2] - \kappa k_2 e^{-\kappa x_2}\sin[(k_2 - 2k_2')x_2] - \kappa(k_2 - 2k_2')e^{-\kappa x_2}\sin(k_2 x_2) - k_2(k_2 - 2k_2')e^{-2\kappa x_2}}{[\kappa^2 + (k_2 - 2k_2')^2](\kappa^2 + k_2^2)[1 + |\vec{k} - \vec{k}'|^2 \lambda^2]^{17/6} (1 + k'^2 \lambda^2)^{17/6}} [(k_1 - k_1')^2 + (k_3 - k_3')^2] k_1' k_3' dA$$

$$+ \int \frac{2k_1^2 k_2^2}{\kappa^2} \frac{2\kappa^2 + \kappa^2 \cos(2k_2 x_2) + 2\kappa k_2 e^{-\kappa x_2}\sin(k_2 x_2) + k_2^2 e^{-2\kappa x_2}}{(\kappa^2 + k_2^2)^2 [1 + |\vec{k} - \vec{k}'|^2 \lambda^2]^{17/6} (1 + k'^2 \lambda^2)^{17/6}} (k_2 - k_2')(k_1 - k_1') k_2' k_1' dA$$

$$- \int \frac{2k_1^2 (k_2 - 2k_2')^2}{\kappa^2} \frac{\kappa^2 \cos[2(k_2 - 2k_2')x_2] - 2\kappa (k_2 - 2k_2')e^{-\kappa x_2}\sin[(k_2 - 2k_2')x_2] - (k_2 - 2k_2')^2 e^{-2\kappa x_2}}{[\kappa^2 + (k_2 - 2k_2')^2]^2 [1 + |\vec{k} - \vec{k}'|^2 \lambda^2]^{17/6} (1 + k'^2 \lambda^2)^{17/6}} (k_2 - k_2')(k_1 - k_1') k_2' k_1' dA$$

$$+ \int \frac{4k_1^2 (k_2 - 2k_2')}{\kappa^2} \frac{\kappa^2 k_2' \cos[2(k_2 - k_2')x_2] - \kappa^2 (k_2 - k_2')\cos(2k_2' x_2) - \kappa k_2^2 e^{-\kappa x_2}\sin[(k_2 - 2k_2')x_2] + \kappa k_2 (k_2 - 2k_2') e^{-\kappa x_2}\sin(k_2 x_2) + k_2^2 (k_2 - 2k_2') e^{-2\kappa x_2}}{[\kappa^2 + (k_2 - 2k_2')^2](\kappa^2 + k_2^2)[1 + |\vec{k} - \vec{k}'|^2 \lambda^2]^{17/6} (1 + k'^2 \lambda^2)^{17/6}} (k_2 - k_2')(k_1 - k_1') k_2' k_1' dA$$

$$+ \int \frac{2k_3^2 k_2^2}{\kappa^2} \frac{2\kappa^2 + \kappa^2 \cos(2k_2 x_2) + 2\kappa k_2 e^{-\kappa x_2}\sin(k_2 x_2) + k_2^2 e^{-2\kappa x_2}}{(\kappa^2 + k_2^2)^2 [1 + |\vec{k} - \vec{k}'|^2 \lambda^2]^{17/6} (1 + k'^2 \lambda^2)^{17/6}} (k_2 - k_2')(k_3 - k_3') k_2' k_3' dA$$

$$- \int \frac{2k_3^2 (k_2 - 2k_2')^2}{\kappa^2} \frac{\kappa^2 \cos[2(k_2 - 2k_2')x_2] - 2\kappa (k_2 - 2k_2')e^{-\kappa x_2}\sin[(k_2 - 2k_2')x_2] - (k_2 - 2k_2')^2 e^{-2\kappa x_2}}{[\kappa^2 + (k_2 - 2k_2')^2]^2 [1 + |\vec{k} - \vec{k}'|^2 \lambda^2]^{17/6} (1 + k'^2 \lambda^2)^{17/6}} (k_2 - k_2')(k_3 - k_3') k_2' k_3' dA$$

$$+ \int \frac{4k_3^2 (k_2 - 2k_2')}{\kappa^2} \frac{\kappa^2 k_2' \cos[2(k_2 - k_2')x_2] - \kappa^2 (k_2 - k_2')\cos(2k_2' x_2) - \kappa k_2^2 e^{-\kappa x_2}\sin[(k_2 - 2k_2')x_2] + \kappa k_2 (k_2 - 2k_2') e^{-\kappa x_2}\sin(k_2 x_2) + k_2^2 (k_2 - 2k_2') e^{-2\kappa x_2}}{[\kappa^2 + (k_2 - 2k_2')^2](\kappa^2 + k_2^2)[1 + |\vec{k} - \vec{k}'|^2 \lambda^2]^{17/6} (1 + k'^2 \lambda^2)^{17/6}} (k_2 - k_2')(k_3 - k_3') k_2' k_3' dA$$

$$+ \int \frac{2k_1 k_3 k_2^2}{\kappa^2} \frac{2\kappa^2 + \kappa^2 \cos(2k_2 x_2) + 2\kappa k_2 e^{-\kappa x_2}\sin(k_2 x_2) + k_2^2 e^{-2\kappa x_2}}{(\kappa^2 + k_2^2)^2 [1 + |\vec{k} - \vec{k}'|^2 \lambda^2]^{17/6} (1 + k'^2 \lambda^2)^{17/6}} (k_2 - k_2')(k_1 - k_1') k_2' k_3' dA$$

$$- \int \frac{2k_1 k_3 (k_2 - 2k_2')^2}{\kappa^2} \frac{\kappa^2 \cos[2(k_2 - 2k_2')x_2] - 2\kappa (k_2 - 2k_2')e^{-\kappa x_2}\sin[(k_2 - 2k_2')x_2] - (k_2 - 2k_2')^2 e^{-2\kappa x_2}}{[\kappa^2 + (k_2 - 2k_2')^2]^2 [1 + |\vec{k} - \vec{k}'|^2 \lambda^2]^{17/6} (1 + k'^2 \lambda^2)^{17/6}} (k_2 - k_2')(k_1 - k_1') k_2' k_3' dA$$

$$+ \int \frac{4k_1 k_3 (k_2 - 2k_2')}{\kappa^2} \frac{\kappa^2 k_2' \cos[2(k_2 - k_2')x_2] - \kappa^2 (k_2 - k_2')\cos(2k_2' x_2) - \kappa k_2^2 e^{-\kappa x_2}\sin[(k_2 - 2k_2')x_2] - \kappa k_2 (k_2 - 2k_2') e^{-\kappa x_2}\sin(k_2 x_2) - k_2^2 (k_2 - 2k_2') e^{-2\kappa x_2}}{[\kappa^2 + (k_2 - 2k_2')^2](\kappa^2 + k_2^2)[1 + |\vec{k} - \vec{k}'|^2 \lambda^2]^{17/6} (1 + k'^2 \lambda^2)^{17/6}} (k_2 - k_2')(k_1 - k_1') k_2' k_3' dA$$

$$+ \int \frac{2k_1 k_3 k_2^2}{\kappa^2} \frac{2\kappa^2 + \kappa^2 \cos(2k_2 x_2) + 2\kappa k_2 e^{-\kappa x_2}\sin(k_2 x_2) + k_2^2 e^{-2\kappa x_2}}{(\kappa^2 + k_2^2)^2 [1 + |\vec{k} - \vec{k}'|^2 \lambda^2]^{17/6} (1 + k'^2 \lambda^2)^{17/6}} (k_2 - k_2')(k_3 - k_3') k_2' k_1' dA$$

$$- \int \frac{2k_1 k_3 (k_2 - 2k_2')^2}{\kappa^2} \frac{\kappa^2 \cos[2(k_2 - 2k_2')x_2] - 2\kappa (k_2 - 2k_2')e^{-\kappa x_2}\sin[(k_2 - 2k_2')x_2] - (k_2 - 2k_2')^2 e^{-2\kappa x_2}}{[\kappa^2 + (k_2 - 2k_2')^2]^2 [1 + |\vec{k} - \vec{k}'|^2 \lambda^2]^{17/6} (1 + k'^2 \lambda^2)^{17/6}} (k_2 - k_2')(k_3 - k_3') k_2' k_1' dA$$

$$+\int \frac{4k_1k_3(k_2-2k_2')\kappa^2 k_2' \cos[2(k_2-k_2')x_2] - \kappa^2(k_2-k_2')\cos(2k_2'x_2) - \kappa k_2^2 e^{\kappa x_2}\sin[(k_2-2k_2')x_2] - \kappa k(k_2-2k_2')e^{\kappa x_2}\sin(k_2x_2) - k_2(k_2-2k_2')\frac{2}{\kappa}e^{\kappa x_2}}{\kappa^2 [\kappa^2+(k_2-2k_2')^2](\kappa^2+k_2^2)[1+|\vec{k}-\vec{k}'|^2\lambda^2]^{17/6}(1+k'^2\lambda^2)^{17/6}}(k_2-k_2')(k_3-k_3')k_2'k_1'dA]$$

$$-[\int \frac{2k_1k_2^3}{\sqrt{2}\kappa^2} \frac{\kappa^2+\kappa^2\cos(2k_2x_2)}{(\kappa^2+k_2^2)^2[1+|\vec{k}-\vec{k}'|^2\lambda^2]^{17/6}(1+k'^2\lambda^2)^{17/6}}[(k_1-k_1')^2+(k_3-k_3')^2]k_2k_1'dA$$

$$+\int \frac{2k_1(k_2-2k_2')^3}{\sqrt{2}\kappa^2} \frac{\kappa^2+\kappa^2\cos[2(k_2-2k_2')x_2]}{[\kappa^2+(k_2-2k_2')^2]^2[1+|\vec{k}-\vec{k}'|^2\lambda^2]^{17/6}(1+k'^2\lambda^2)^{17/6}}[(k_1-k_1')^2+(k_3-k_3')^2]k_2'k_1'dA$$

$$-\int \frac{4k_1k_2k_2'(k_2-2k_2')}{\sqrt{2}\kappa^2} \frac{\kappa^2\cos(2k_2'x_2)+\kappa^2\cos[2(k_2-k_2')x_2]}{[\kappa^2+(k_2-2k_2')^2](\kappa^2+k_2^2)[1+|\vec{k}-\vec{k}'|^2\lambda^2]^{17/6}(1+k'^2\lambda^2)^{17/6}}[(k_1-k_1')^2+(k_3-k_3')^2]k_2'k_1'dA$$

$$-\int \frac{2k_1k_2^3}{\sqrt{2}\kappa^2} \frac{\kappa^2+\kappa^2\cos(2k_2x_2)}{(\kappa^2+k_2^2)^2[1+|\vec{k}-\vec{k}'|^2\lambda^2]^{17/6}(1+k'^2\lambda^2)^{17/6}}(k_2-k_2')(k_1-k_1')(k_1'^2+k_3'^2)dA$$

$$-\int \frac{2k_1(k_2-2k_2')^3}{\sqrt{2}\kappa^2} \frac{\kappa^2+\kappa^2\cos[2(k_2-2k_2')x_2]}{[\kappa^2+(k_2-2k_2')^2]^2[1+|\vec{k}-\vec{k}'|^2\lambda^2]^{17/6}(1+k'^2\lambda^2)^{17/6}}(k_2-k_2')(k_1-k_1')(k_1'^2+k_3'^2)dA$$

$$+\int \frac{4k_1k_2(k_2-2k_2')(k_2-k_2')}{\sqrt{2}\kappa^2} \frac{\kappa^2\cos(2k_2'x_2)+\kappa^2\cos[2(k_2-k_2')x_2]}{[\kappa^2+(k_2-2k_2')^2](\kappa^2+k_2^2)[1+|\vec{k}-\vec{k}'|^2\lambda^2]^{17/6}(1+k'^2\lambda^2)^{17/6}}(k_2-k_2')(k_1-k_1')(k_1'^2+k_3'^2)dA$$

$$-\int \frac{2k_3k_2^3}{\sqrt{2}\kappa^2} \frac{\kappa^2+\kappa^2\cos(2k_2x_2)}{(\kappa^2+k_2^2)^2[1+|\vec{k}-\vec{k}'|^2\lambda^2]^{17/6}(1+k'^2\lambda^2)^{17/6}}[(k_1-k_1')^2+(k_3-k_3')^2]k_2'k_3'dA$$

$$+\int \frac{2k_3(k_2-2k_2')^3}{\sqrt{2}\kappa^2} \frac{\kappa^2+\kappa^2\cos[2(k_2-2k_2')x_2]}{[\kappa^2+(k_2-2k_2')^2]^2[1+|\vec{k}-\vec{k}'|^2\lambda^2]^{17/6}(1+k'^2\lambda^2)^{17/6}}[(k_1-k_1')^2+(k_3-k_3')^2]k_2'k_3'dA$$

$$-\int \frac{4k_3k_2k_2'(k_2-2k_2')}{\sqrt{2}\kappa^2} \frac{\kappa^2\cos(2k_2'x_2)+\kappa^2\cos[2(k_2-k_2')x_2]}{[\kappa^2+(k_2-2k_2')^2](\kappa^2+k_2^2)[1+|\vec{k}-\vec{k}'|^2\lambda^2]^{17/6}(1+k'^2\lambda^2)^{17/6}}[(k_1-k_1')^2+(k_3-k_3')^2]k_2'k_3'dA$$

$$-\int \frac{2k_3k_2^3}{\sqrt{2}\kappa^2} \frac{\kappa^2+\kappa^2\cos(2k_2x_2)}{(\kappa^2+k_2^2)^2[1+|\vec{k}-\vec{k}'|^2\lambda^2]^{17/6}(1+k'^2\lambda^2)^{17/6}}(k_2-k_2')(k_3-k_3')(k_1'^2+k_3'^2)dA$$

$$-\int \frac{2k_3(k_2-2k_2')^3}{\sqrt{2}\kappa^2} \frac{\kappa^2+\kappa^2\cos[2(k_2-2k_2')x_2]}{[\kappa^2+(k_2-2k_2')^2]^2[1+|\vec{k}-\vec{k}'|^2\lambda^2]^{17/6}(1+k'^2\lambda^2)^{17/6}}(k_2-k_2')(k_3-k_3')(k_1'^2+k_3'^2)dA$$

$$+\int \frac{4k_3k_2(k_2-2k_2')(k_2-k_2')}{\sqrt{2}\kappa^2} \frac{\kappa^2\cos(2k_2'x_2)+\kappa^2\cos[2(k_2-k_2')x_2]}{[\kappa^2+(k_2-2k_2')^2](\kappa^2+k_2^2)[1+|\vec{k}-\vec{k}'|^2\lambda^2]^{17/6}(1+k'^2\lambda^2)^{17/6}}(k_2-k_2')(k_3-k_3')(k_1'^2+k_3'^2)dA]$$

$$+[\int \frac{k_2^4}{\kappa^2} \frac{\kappa^2+\kappa^2\cos(2k_2x_2)}{(\kappa^2+k_2^2)^2[1+|\vec{k}-\vec{k}'|^2\lambda^2]^{17/6}(1+k'^2\lambda^2)^{17/6}}[(k_1-k_1')^2+(k_3-k_3')^2](k_1'^2+k_3'^2)dA$$

$$+\int \frac{(k_2-2k_2')^4}{\kappa^2} \frac{\kappa^2+\kappa^2\cos[(2k_2-4k_2')x_2]}{[\kappa^2+(k_2-2k_2')^2]^2[1+|\vec{k}-\vec{k}'|^2\lambda^2]^{17/6}(1+k'^2\lambda^2)^{17/6}}[(k_1-k_1')^2+(k_3-k_3')^2](k_1'^2+k_3'^2)dA$$

$$-\int \frac{2k_2^2(k_2-2k_2')^2}{\kappa^2} \frac{\kappa^2\cos(2k_2'x_2)+\kappa^2\cos[(2k_2-2k_2')x_2]}{[\kappa^2+(k_2-2k_2')^2](\kappa^2+k_2^2)[1+|\vec{k}-\vec{k}'|^2\lambda^2]^{17/6}(1+k'^2\lambda^2)^{17/6}}[(k_1-k_1')^2+(k_3-k_3')^2](k_1'^2+k_3'^2)dA]\}$$

APPENDIX D. Form of $|p(k_1)|^2$ used for calculations using the same method but with homogeneous isotropic turbulence input in Eq. 2

For calculations using the same method but with homogeneous isotropic turbulence input, the pressure fluctuation should have no height dependence. The boundary is removed from the integration. There is

$$p(\vec{\kappa},\omega) = \frac{1}{2}\kappa^{-1}\int_{-\infty}^{\infty} e^{-\kappa|x_2'|}T(x_2',\vec{\kappa},\omega)dx_2', \quad (D-1)$$

where $T(x_2,\vec{\kappa},\omega)$ is the Fourier transform of the source function $T(\vec{x},t)$ with respect to $x_1$, $x_3$, and $t$.

The power spectrum form of the pressure transform normalized to unit area and time is

$$|p(\vec{\kappa},\omega)|^2$$
$$=\frac{1}{4}(2\pi)^{-\frac{3}{2}}\kappa^{-2}[\int_0^\infty\int_0^\infty e^{-\kappa x_2'-\kappa x_2''}S(x_2'',x_2',\vec{\kappa},\omega)dx_2'dx_2''+\int_{-\infty}^0\int_0^\infty e^{-\kappa x_2'+\kappa x_2''}S(x_2'',x_2',\vec{\kappa},\omega)dx_2'dx_2'' \quad\quad \text{(D-2)}$$
$$+\int_0^\infty\int_{-\infty}^0 e^{\kappa x_2'-\kappa x_2''}S(x_2'',x_2',\vec{\kappa},\omega)dx_2'dx_2''+\int_{-\infty}^0\int_{-\infty}^0 e^{\kappa x_2'+\kappa x_2''}S(x_2'',x_2',\vec{\kappa},\omega)dx_2'dx_2''].$$

$|p(k_1)|^2$ for calculations using the same method but with homogeneous isotropic turbulence input has the form below. Note that $\int dA$ refers to $\int_{-\infty}^\infty\int_{-\infty}^\infty\int_{-\infty}^\infty\int_{-\infty}^\infty\int_{-\infty}^\infty dk_1'dk_2'dk_3'dk_2dk_3$ for simplicity.

$$|p(k_1)|^2$$
$$=(2\pi)^{-2}(\frac{55C}{18})^2\rho^2\lambda^8\{[\int k_1^4\frac{1}{(\kappa^2+k_2^2)^2[1+|\vec{k}-\vec{k}'|^2\lambda^2]^{17/6}(1+k'^2\lambda^2)^{17/6}}[(k_2-k_2')^2+(k_3-k_3')^2](k_2'^2+k_3'^2)dA$$
$$+\int k_3^4\frac{1}{(\kappa^2+k_2^2)^2[1+|\vec{k}-\vec{k}'|^2\lambda^2]^{17/6}(1+k'^2\lambda^2)^{17/6}}[(k_1-k_1')^2+(k_2-k_2')^2](k_1'^2+k_2'^2)dA$$
$$+\int k_1^2k_3^2\frac{1}{(\kappa^2+k_2^2)^2[1+|\vec{k}-\vec{k}'|^2\lambda^2]^{17/6}(1+k'^2\lambda^2)^{17/6}}[(k_2-k_2')^2+(k_3-k_3')^2](k_1'^2+k_2'^2)dA$$
$$+\int k_1^2k_3^2\frac{1}{(\kappa^2+k_2^2)^2[1+|\vec{k}-\vec{k}'|^2\lambda^2]^{17/6}(1+k'^2\lambda^2)^{17/6}}[(k_1-k_1')^2+(k_2-k_2')^2](k_2'^2+k_3'^2)dA$$
$$+\int 2k_1^3k_3\frac{1}{(\kappa^2+k_2^2)^2[1+|\vec{k}-\vec{k}'|^2\lambda^2]^{17/6}(1+k'^2\lambda^2)^{17/6}}[(k_2-k_2')^2+(k_3-k_3')^2](-k_1'k_3')dA$$
$$+\int 2k_1^3k_3\frac{1}{(\kappa^2+k_2^2)^2[1+|\vec{k}-\vec{k}'|^2\lambda^2]^{17/6}(1+k'^2\lambda^2)^{17/6}}[-(k_1-k_1')(k_3-k_3')](k_2'^2+k_3'^2)dA$$
$$+\int 2k_1k_3^3\frac{1}{(\kappa^2+k_2^2)^2[1+|\vec{k}-\vec{k}'|^2\lambda^2]^{17/6}(1+k'^2\lambda^2)^{17/6}}[(k_1-k_1')^2+(k_2-k_2')^2](-k_1'k_3')dA$$
$$+\int 2k_1k_3^3\frac{1}{(\kappa^2+k_2^2)^2[1+|\vec{k}-\vec{k}'|^2\lambda^2]^{17/6}(1+k'^2\lambda^2)^{17/6}}[-(k_1-k_1')(k_3-k_3')](k_1'^2+k_2'^2)dA \quad\quad \text{(D-3)}$$
$$+\int 4k_1^2k_3^2\frac{1}{(\kappa^2+k_2^2)^2[1+|\vec{k}-\vec{k}'|^2\lambda^2]^{17/6}(1+k'^2\lambda^2)^{17/6}}[-(k_1-k_1')(k_3-k_3')](-k_1'k_3')dA]$$
$$+[\int 2k_1^2k_3k_2\frac{1}{(\kappa^2+k_2^2)^2[1+|\vec{k}-\vec{k}'|^2\lambda^2]^{17/6}(1+k'^2\lambda^2)^{17/6}}[(k_2-k_2')^2+(k_3-k_3')^2](-k_2'k_3')dA$$
$$+\int 2k_1^2k_3k_2\frac{1}{(\kappa^2+k_2^2)^2[1+|\vec{k}-\vec{k}'|^2\lambda^2]^{17/6}(1+k'^2\lambda^2)^{17/6}}[-(k_2-k_2')(k_3-k_3')](k_2'^2+k_3'^2)dA$$
$$+\int 2k_3^3k_2\frac{1}{(\kappa^2+k_2^2)^2[1+|\vec{k}-\vec{k}'|^2\lambda^2]^{17/6}(1+k'^2\lambda^2)^{17/6}}[(k_1-k_1')^2+(k_2-k_2')^2](-k_2'k_3')dA$$
$$+\int 2k_3^3k_2\frac{1}{(\kappa^2+k_2^2)^2[1+|\vec{k}-\vec{k}'|^2\lambda^2]^{17/6}(1+k'^2\lambda^2)^{17/6}}[-(k_2-k_2')(k_3-k_3')](k_1'^2+k_2'^2)dA$$
$$+\int 2k_1^3k_2\frac{1}{(\kappa^2+k_2^2)^2[1+|\vec{k}-\vec{k}'|^2\lambda^2]^{17/6}(1+k'^2\lambda^2)^{17/6}}[(k_2-k_2')^2+(k_3-k_3')^2](-k_2'k_1')dA$$
$$+\int 2k_1^3k_2\frac{1}{(\kappa^2+k_2^2)^2[1+|\vec{k}-\vec{k}'|^2\lambda^2]^{17/6}(1+k'^2\lambda^2)^{17/6}}[-(k_2-k_2')(k_1-k_1')](k_2'^2+k_3'^2)\,dA$$

$$+\int 2k_1k_3^2k_2 \frac{1}{(\kappa^2+k_2^2)^2[1+|\vec{k}-\vec{k}'|^2\lambda^2]^{17/6}(1+k'^2\lambda^2)^{17/6}}[(k_1-k_1')^2+(k_2-k_2')^2](-k_2'k_1')dA$$

$$+\int 2k_1k_3^2k_2 \frac{1}{(\kappa^2+k_2^2)^2[1+|\vec{k}-\vec{k}'|^2\lambda^2]^{17/6}(1+k'^2\lambda^2)^{17/6}}[-(k_2-k_2')(k_1-k_1')](k_1'^2+k_2'^2)dA$$

$$+\int 4k_1^2k_3k_2 \frac{1}{(\kappa^2+k_2^2)^2[1+|\vec{k}-\vec{k}'|^2\lambda^2]^{17/6}(1+k'^2\lambda^2)^{17/6}}[-(k_1-k_1')(k_3-k_3')](-k_2'k_1')dA$$

$$+\int 4k_1^2k_3k_2 \frac{1}{(\kappa^2+k_2^2)^2[1+|\vec{k}-\vec{k}'|^2\lambda^2]^{17/6}(1+k'^2\lambda^2)^{17/6}}[-(k_2-k_2')(k_1-k_1')](-k_1'k_3')dA$$

$$+\int 4k_1k_3^2k_2 \frac{1}{(\kappa^2+k_2^2)^2[1+|\vec{k}-\vec{k}'|^2\lambda^2]^{17/6}(1+k'^2\lambda^2)^{17/6}}[-(k_1-k_1')(k_3-k_3')](-k_2'k_3')dA$$

$$+\int 4k_1k_3^2k_2 \frac{1}{(\kappa^2+k_2^2)^2[1+|\vec{k}-\vec{k}'|^2\lambda^2]^{17/6}(1+k'^2\lambda^2)^{17/6}}[-(k_2-k_2')(k_3-k_3')](-k_1'k_3')dA]$$

$$+[\int k_1^2k_2^2 \frac{1}{(\kappa^2+k_2^2)^2[1+|\vec{k}-\vec{k}'|^2\lambda^2]^{17/6}(1+k'^2\lambda^2)^{17/6}}[(k_2-k_2')^2+(k_3-k_3')^2](k_1'^2+k_3'^2)dA$$

$$+\int k_1^2k_2^2 \frac{1}{(\kappa^2+k_2^2)^2[1+|\vec{k}-\vec{k}'|^2\lambda^2]^{17/6}(1+k'^2\lambda^2)^{17/6}}[(k_1-k_1')^2+(k_3-k_3')^2](k_2'^2+k_3'^2)dA$$

$$+\int k_3^2k_2^2 \frac{1}{(\kappa^2+k_2^2)^2[1+|\vec{k}-\vec{k}'|^2\lambda^2]^{17/6}(1+k'^2\lambda^2)^{17/6}}[(k_1-k_1')^2+(k_2-k_2')^2](k_1'^2+k_3'^2)dA$$

$$+\int k_3^2k_2^2 \frac{1}{(\kappa^2+k_2^2)^2[1+|\vec{k}-\vec{k}'|^2\lambda^2]^{17/6}(1+k'^2\lambda^2)^{17/6}}[(k_1-k_1')^2+(k_3-k_3')^2](k_1'^2+k_2'^2)dA$$

$$+\int 2k_1k_3k_2^2 \frac{1}{(\kappa^2+k_2^2)^2[1+|\vec{k}-\vec{k}'|^2\lambda^2]^{17/6}(1+k'^2\lambda^2)^{17/6}}[-(k_1-k_1')(k_3-k_3')](k_1'^2+k_3'^2)dA$$

$$+\int 2k_1k_3k_2^2 \frac{1}{(\kappa^2+k_2^2)^2[1+|\vec{k}-\vec{k}'|^2\lambda^2]^{17/6}(1+k'^2\lambda^2)^{17/6}}[(k_1-k_1')^2+(k_3-k_3')^2](-k_1'k_3')dA$$

$$+\int 4k_1^2k_2^2 \frac{1}{(\kappa^2+k_2^2)^2[1+|\vec{k}-\vec{k}'|^2\lambda^2]^{17/6}(1+k'^2\lambda^2)^{17/6}}[-(k_2-k_2')(k_1-k_1')](-k_2'k_1')dA$$

$$+\int 4k_3^2k_2^2 \frac{1}{(\kappa^2+k_2^2)^2[1+|\vec{k}-\vec{k}'|^2\lambda^2]^{17/6}(1+k'^2\lambda^2)^{17/6}}[-(k_2-k_2')(k_3-k_3')](-k_2'k_3')dA$$

$$+\int 4k_1k_3k_2^2 \frac{1}{(\kappa^2+k_2^2)^2[1+|\vec{k}-\vec{k}'|^2\lambda^2]^{17/6}(1+k'^2\lambda^2)^{17/6}}[-(k_2-k_2')(k_1-k_1')](-k_2'k_3')dA$$

$$+\int 4k_1k_3k_2^2 \frac{1}{(\kappa^2+k_2^2)^2[1+|\vec{k}-\vec{k}'|^2\lambda^2]^{17/6}(1+k'^2\lambda^2)^{17/6}}[-(k_2-k_2')(k_3-k_3')](-k_2'k_1')dA]$$

$$+[\int 2k_1k_2^3 \frac{1}{(\kappa^2+k_2^2)^2[1+|\vec{k}-\vec{k}'|^2\lambda^2]^{17/6}(1+k'^2\lambda^2)^{17/6}}[(k_1-k_1')^2+(k_3-k_3')^2](-k_2'k_1')dA$$

$$+\int 2k_1k_2^3 \frac{1}{(\kappa^2+k_2^2)^2[1+|\vec{k}-\vec{k}'|^2\lambda^2]^{17/6}(1+k'^2\lambda^2)^{17/6}}[-(k_2-k_2')(k_1-k_1')](k_1'^2+k_3'^2)dA$$

$$+\int 2k_3k_2^3 \frac{1}{(\kappa^2+k_2^2)^2[1+|\vec{k}-\vec{k}'|^2\lambda^2]^{17/6}(1+k'^2\lambda^2)^{17/6}}[(k_1-k_1')^2+(k_3-k_3')^2](-k_2'k_3')dA$$

$$+\int 2k_3k_2^3 \frac{1}{(\kappa^2+k_2^2)^2[1+|\vec{k}-\vec{k}'|^2\lambda^2]^{17/6}(1+k'^2\lambda^2)^{17/6}}[-(k_2-k_2')(k_3-k_3')](k_1'^2+k_3'^2)dA]$$

$$+[\int k_2^4 \frac{1}{(\kappa^2+k_2^2)^2[1+|\vec{k}-\vec{k}'|^2\lambda^2]^{17/6}(1+k'^2\lambda^2)^{17/6}}[(k_1-k_1')^2+(k_3-k_3')^2](k_1'^2+k_3'^2)dA]\}$$

APPENDIX E. The derivation process for the one dimensional turbulence-turbulence interaction pressure spectrum from Eq. 17 for homogeneous isotropic turbulence.

For homogeneous isotropic turbulence, the one dimensional turbulence-turbulence interaction pressure spectrum $F^1_{ppt}(k_1)$ can be related to $F^t_{p,p}(\vec{k})$ as [4,6]:

$$F^1_{ppt}(k_1) = 2\pi \int_{k_1}^{\infty} k F^t_{p,p}(\vec{k}) dk, \tag{E-1}$$

and since [4]

$$F_{LL,LL} = F_{11,11}(k,0,0) \tag{E-2}$$

we obtain

$$F^1_{ppt}(k_1) = \rho^2 2\pi \int_{k_1}^{\infty} k F_{11,11}(k,0,0) dk, \tag{E-3}$$

where the subscript 1 indicates that $F_{11,11}$ is expressed in the regular Cartesian coordinate system $Ox_1x_2x_3$. The integration in the equation above can be transformed to the Cartesian coordinate:

$$2\pi \int_{k_1}^{\infty} k F(\vec{k}) dk = \int_{-\infty}^{\infty}\int_{-\infty}^{\infty} F(\vec{k}) dk_2 dk_3, \tag{E-4}$$

so

$$F^1_{ppt}(k_1) = \rho^2 \int_{-\infty}^{\infty}\int_{-\infty}^{\infty} F_{11,11}(k,0,0) dk_2 dk_3. \tag{E-5}$$

Because fourth-order moments of the turbulence and second-order moments satisfy joint Gaussianity assumptions, there is [4,5]

$$F_{\alpha\beta,\mu\nu}(\vec{k}) = \int F_{\alpha,\mu}(\vec{k}-\vec{k}')F_{\beta,\nu}(\vec{k}')d^3\vec{k}' + \int F_{\alpha,\nu}(\vec{k}-\vec{k}')F_{\beta,\mu}(\vec{k}')d^3\vec{k}', \tag{E-6}$$

combining with Eq. 14 and Eq. 16, Eq. E-5 becomes

$$F^1_{ppt}(k_1) = (2\pi)^{-1}(\frac{55C}{18})^2 \rho^2 \lambda^8 \int \frac{k(k_2'^2+k_3'^2)(k_2'^2+k_3'^2)}{[1+[(k-k_1')^2+k_2'^2+k_3'^2]\lambda^2]^{17/6}[1+(k_1'^2+k_2'^2+k_3'^2)\lambda^2]^{17/6}} dB. \tag{E-7}$$

Note that $\int dB$ refers to $\int_{k_1}^{\infty}\int_{-\infty}^{\infty}\int_{-\infty}^{\infty}\int_{-\infty}^{\infty} dk_1' dk_2' dk_3' dk$ for simplicity. To convert double-sided spectrum to one-sided spectrum with only positive $k_1$, a factor of two is introduced.

The one dimensional turbulence-turbulence interaction pressure spectral contribution from the $F_{11}F_{11}$ term for isotropic turbulence in Appendix D is

$$|p(k_1)|^2_{11,11} = (2\pi)^{-2}(\frac{55C}{18})^2 \rho^2 \lambda^8 \int k_1^4 \frac{1}{(\kappa^2+k_2^2)^2[1+|\vec{k}-\vec{k}'|^2\lambda^2]^{17/6}(1+k'^2\lambda^2)^{17/6}}[(k_2-k_2')^2+(k_3-k_3')^2](k_2'^2+k_3'^2) dA. \tag{E-8}$$

Comparing Eq. E-7 and Eq. E-8, and taking Eq. E-4 into consideration, the difference between Eq. E-7 and Eq. E-8 exists partly in that $\vec{k} = (k,0,0)$ in Eq. E-7, and partly in that $\frac{k_1^4}{k^4}$ being introduced in Eq. E-8. The relation [4]

$$\frac{1}{\rho^2}F^t_{p,p}(\vec{k}) = \frac{k_i k_j k_l k_m}{k^4}F_{ij,lm}(\vec{k}) \tag{E-9}$$

justifies the introduction of $\frac{k_1^4}{k^4}$ in Eq. E-8 and Eq. D-3.